\documentclass[11pt,letterpaper]{article}
\usepackage{mathtools}
\usepackage{axodraw2}
\usepackage{tikz}
\usepackage{tikz-feynman}
\usepackage{amscd}
\usepackage{amsmath}
\usepackage{slashed}
\usepackage{gensymb}
\usepackage{bm}
\usepackage{jheppub} 

\newcommand{\bsll}{$b \rightarrow s \ell^+ \ell^-$}
\newcommand{\bsmm}{$b \rightarrow s \mu^+ \mu^-$}
\newcommand{\bsee}{$b \rightarrow s e^+ e^-$}

\newcommand{\smelli}{{\tt smelli2.3.2}}
\newcommand{\flavio}{{\tt flavio2.3.3}}
\newcommand{\wilson}{{\tt wilson2.3.2}}

\newcommand{\flavioA}{{\tt flavio2.3.3}}

\newcommand{\smelliB}{{\tt smelli2.4.2}}
\newcommand{\flavioB}{{\tt flavio2.6.2}}
\newcommand{\wilsonB}{{\tt wilson2.5.2}}

\title{\centering
  Plan B:\\
  New {\boldmath $Z^\prime$} models for {\boldmath \bsll}\ anomalies}

\author[a]{Ben Allanach}
\affiliation[a]{DAMTP, University of Cambridge, Wilberforce Road, Cambridge, 
  CB3 0WA, United Kingdom}
\emailAdd{B.C.Allanach@damtp.cam.ac.uk}

\author[b,1]{and Anna Mullin\note{Corresponding author.}}
\affiliation[b]{Department of Physics, Cavendish Laboratory, JJ Thomson Avenue,
  Cambridge, CB3 0HE, United Kingdom}
\emailAdd{ajm351@cam.ac.uk}

\abstract{Measurements of \bsmm\ transitions indicate that there may be
  a new physics field coupling to di-muon pairs associated with the
$b$ to $s$ flavour transition. Including the 2022
LHCb reanalysis of $R_K$ and 
$R_{K^\ast}$, one infers that there may also be associated new
physics in \bsee\ transitions.
Here, we examine the extent of the statistical
preference for $Z^\prime$ models coupling to 
di-electron pairs taking into account the relevant constraints,
in particular from experiments at LEP-2.
We identify an anomaly-free set of models
which interpolates between the $Z^\prime$ not coupling to electrons at all, to
one in which there is
an equal $Z^\prime$ coupling to muons and electrons (but where in all
models in the set, the $Z^\prime$ boson can mediate \bsmm\ transitions).
A $3B_3-L_e-2L_\mu$ model provides a close-to-optimal fit 
to the pertinent measurements along the line of interpolation.
We have (re-)calculated predictions for the relevant LEP-2 observables
in terms of dimension-6 SMEFT operators
and put them into the {\tt flavio} computer program, so that
they are available for global fits.}   


\begin{document} 

\maketitle
\flushbottom

\section{Introduction \label{sec:intro}}

Various measurements of $B-$meson decays at LHC experiments are in
tension with 
Standard Model (SM) predictions, particularly when the final state includes a
di-muon pair. For example the CMS, ATLAS and LHCb combined~\cite{ATLAS:2018cur,CMS:2022dbz,LHCb:2017rmj}
$CP-$untagged, time integrated branching ratio of $B_s$ decaying to di-muon
pairs $\overline{BR}(B_s\rightarrow \mu^+\mu^-)$~\cite{ATLAS:2018cur,CMS-PAS-BPH-21-006,LHCb:2017rmj} has a 1.6$\sigma$ tension~\cite{Allanach:2022iod} with SM
predictions. 
Furthermore, measurements in various di-muon invariant mass-squared ($q^2$)
bins of 
$BR(B_s \rightarrow \phi \mu^+ \mu^-)$ are up to $4\sigma$ smaller than the SM
predictions~\cite{LHCb:2021zwz,CDF:2012qwd}. Some angular distributions in $B \rightarrow K^\ast \mu^+ \mu^-$
decays have been measured by LHC
experiments~\cite{LHCb:2013ghj,LHCb:2015svh,ATLAS:2018gqc,CMS:2017rzx,CMS:2015bcy,Bobeth:2017vxj}
to be several $\sigma$ short of state-of-the-art SM
predictions and
the same can be said of $BR(B \rightarrow K^\ast \mu^+
\mu^-)$~\cite{Parrott:2022zte}. 
We call the aforementioned tensions the neutral current $b\rightarrow s
\mu^+\mu^-$ anomalies.
It is tempting to suppose that the tensions could be explained by
unaccounted-for new physics.
It has been shown in 
fits~\cite{Alguero:2023jeh,Wen:2023pfq} to measurements that the weak
effective theory (WET)
operators  
\begin{equation}
  {\mathcal L} = \ldots + N \left( C_9^{(\mu)} (\bar b \gamma^\alpha P_L s) (\bar \mu
  \gamma_\alpha \mu)  + C_{10}^{(\mu)}(\bar b \gamma^\alpha P_L s) (\bar \mu
  \gamma_\alpha \gamma_5 \mu)  + H.c.\right), \label{c9mu}
\end{equation}
parameterising the effects of new physics states, significantly improve the
situation. 
Such beyond-the-SM operators may be
generated by integrating out putative heavy new physics states. Here, $b$ is
the bottom quark field, $\mu$ the muon field and $s$ the strange quark field.
$N:=4 G_F e^2 |V_{ts}| / (16 \pi^2\sqrt{2})$ is a normalising constant, where
$G_F$ is the Fermi decay constant, $e$ the electromagnetic gauge coupling and
$V_{ij}$ the entries of the CKM matrix.

We use the \smelli{}~\cite{Aebischer:2018iyb} computer program to predict the
aforementioned 
\bsmm\ anomalies. \smelli\ puts them in the `quarks' category of
observable. For these, there is some 
debate about the most accurate predictions and the
size of the associated theoretical
uncertainties
although many estimates (e.g.\ \cite{Gubernari:2022hxn,Gubernari:2023puw})
predict that the 
theoretical uncertainties alone cannot explain the
\bsmm\ anomalies\footnote{Some 
estimates in Ref.~\cite{Ciuchini:2022wbq} fit an unidentified non-perturbative
SM contribution
that mimics a $q^2-$dependent lepton-family universal $C_9$ in tandem with the
new physics operators.
As argued in Ref.~\cite{Isidori:2023unk}, a similar non-perturbative effect
cannot explain the 2$\sigma$ deficit in the
$BR(B \rightarrow X_s 
\mu^+ \mu^-)$ high $q^2-$bin, which is compatible with the low $q^2-$deficits. 
The \bsmm\ anomalies persist when one uses ratios of observables including $\Delta
M_{s,d}$, $\epsilon_K$, $S_{\psi K_S}$ to cancel their dependence on CKM
matrix elements~\cite{Buras:2022wpw,Buras:2022qip}, although throughout the
present paper, new physics contributions to CKM matrix elements are predicted
to be negligible.}.

Contrary to the \bsmm\ anomalies, a 2022 LHCb reanalysis~\cite{LHCb:2022qnv}
holds that measurements of 
\begin{equation}
  R_{A}(q^2_\text{min},\ q^2_\text{max}) :=
  \frac{\int_{q^2_\text{min}}^{q^2_\text{max}} dq^2
    BR(B \rightarrow A\mu^+\mu^-(q^2) )}
  {\int_{q^2_\text{min}}^{q^2_\text{max}} dq^2
  BR(B \rightarrow A e^+e^-(q^2) )}, \label{Rdef}
\end{equation}
are broadly \emph{compatible}
with SM predictions for $A \in \{K, K^\ast\}$, within
  uncertainties\footnote{There are some $1\sigma$ mild tensions, however, for $A={K_S^0}$ and
$A=K^{\ast\pm}$~\cite{LHCb:2021lvy}.}.  
Such ratios are commonly called lepton flavour
universality (LFU) variables. 
Since we entertain the possibility that the
\bsmm\ anomalies may be pointing to some new physics state coupling
to muons (and $\bar b s$ quarks) the reanalysis suggests that there could
also be a new physics contribution from
\begin{equation}
  {\mathcal L} = \ldots + N \left( C_9^{(e)} (\bar b \gamma^\alpha P_L s) (\bar e
  \gamma_\alpha e)  + C_{10}^{(e)}(\bar b \gamma^\alpha P_L s) (\bar e
  \gamma_\alpha \gamma_5 e)  + H.c.\right). \label{c9e}
\end{equation}
This possibility has already been partially addressed in
Ref.~\cite{Alguero:2023jeh}, where 
constraints on the $C_9^{(e)}-C_9^{(\mu)}$ parameter plane from some different
relevant flavour observables were presented, where
all other new physics Wilson coefficients are null.
It was demonstrated that there is parameter space where the
constraints are compatible with each other at the 95$\%$ confidence level
(CL).
Some cases with other non-zero new physics Wilson operators were also analysed
in Refs.~\cite{Alguero:2023jeh,Wen:2023pfq}.
It was shown in Ref.~\cite{Alguero:2023jeh} that, fitting two dominant
new-physics WET operators to $b\rightarrow s \mu^+\mu^-$ data, $C_9^{(e)}$ and
$C_9^{(\mu)}$ provide the biggest fit improvement upon the SM compared to other
scenarios involving new physics effects with right-handed quark currents
(${C_9^\prime}^{(\mu)}$ and ${C_{10}^\prime}^{(\mu)}$) or other operators.  
It had already been emphasised though
that current data on direct $CP-$violation in $B\rightarrow K \mu\mu$ decays
coupled with measurements of the branching ratio and the 2022 LHCb constraints
upon $R_{K^{(\ast)}}$ still allow significant lepton universality \emph{violation}
between 
$C_{9,10}^{(\mu)}$ and $C_{9,10}^{(e)}$~\cite{Fleischer:2023zeo}. 
We note here that often, the
natural language to describe the interactions of TeV-scale models is the SM
effective field theory (SMEFT)~\cite{Grzadkowski:2010es}, which involves
complete representations of 
the unbroken SM gauge group (e.g.\ $SU(2)_L$ doublets), as opposed to WET,
which is valid below the $W$ boson mass and is therefore 
written in the spontaneously broken phase of the electroweak gauge symmetry.

Within the present paper, we shall only address the \bsmm\ anomalies, not the
charged current anomalies in $b \rightarrow c \ell \bar \nu$ transitions, which currently
display a joint deviation between two particular SM predictions and
measurements~\cite{HFLAV2} at the 3.3$\sigma$ level. Were
this deviation to become definite and 
confirmed, the models and scenarios contained within the present paper would
require significant modification, for example by adding additional charged
gauge fields or leptoquarks, with family-dependent interactions.

In the following section, we shall perform our own fits including the new
physics operators in (\ref{c9mu}) and (\ref{c9e}) in order to check
the 
compatibility of some of the results of Ref.~\cite{Alguero:2023jeh} with the
different 
theoretical calculation of \smelli{}. Then, in
\S\ref{sec:models}, we examine $Z^\prime$ models that are capable of
predicting them. Some of the other operators induced yield
a change to di-lepton production cross-section
measurements at experiments at the LEP-2 collider, which we recalculate in
\S\ref{sec:LEP}. Using these constraints, we examine the fits to our set of
models in \S\ref{sec:fits}, quantifying the extent to which a non-zero
coupling of the 
$Z^\prime$ to di-electron pairs is preferred.
One particular model based on $U(1)_{3B_3-L_e-2L_\mu}$ is singled out as being
close-to-optimal whilst simultaneously having relatively low $U(1)$ charges
for the fermionic fields. 
Parameter space constraints are
presented. We summarise and conclude in \S\ref{sec:conc}.

\section{SMEFT operator fit \label{sec:SMEFTop}}

Introducing operators that couple di-electron and di-muon pairs with new
physics appears to significantly improve fits to recent measurements. In
Section~\ref{sec:fits} we investigate the best fit for $Z^{\prime }$ models
described in Section~\ref{sec:models}, but to inform our choice of
model we first understand the phenomenological effects of adding
only four non-zero Wilson coefficients (WCs): $C_9^{(e)}$, $C_9^{(\mu)}$,
$C_{10}^{(e)}$ and 
$C_{10}^{(\mu)}$. 
Note that in particular we do not consider possible contributions to isospin
triplet operators (which may induce changes to $b \rightarrow c \ell \bar \nu$
transitions), nor do we consider purely right-handed quark current
contributions (as mentioned in \S\ref{sec:intro}, these can ameliorate the
fit to neutral current $b-$anomalies but they do not provide the best fit
improvement, at least in simplified set-ups). 
Within the restricted set of operators that we consider -- which are
generated by the $Z^\prime$ models we consider later -- we
check how $b\rightarrow s \mu^+ \mu^-$ measurements and LFU observables
(especially the $R_K$ and $R_{K^*}$ ratios from the 2022 LHCb
reanalysis~\cite{LHCb:2022qnv}) affect the statistical preference for new
physics that couples to di-electron pairs.     

We place constraints and perform global fits in the parameter plane $C_9^{(e)}
- C_9^{(\mu )}$ similar to Ref.~\cite{Alguero:2023jeh}. Our evaluation focuses
on four cases which encompass combinations of left-handed ($C_9^a=-C_{10}^a$ for
$a \in \{e,\mu\}$) and vector-like ($C_{10}^a=0$) couplings of new physics to
di-muon pairs and/or
di-electron pairs through appropriate selection of our chosen WCs.
We shall take two-dimensional slices through the four-dimensional parameter
space using combinations of these couplings.

The WCs introduced above belong to the WET, whereas
we give inputs to 
\smelli\ belonging to the Standard Model effective field theory
(SMEFT) WCs. The SMEFT provides a framework for describing new physics
contributions at energies much larger than the electroweak scale. We match between the WET
Hamiltonian and the SMEFT operators as described
in Ref.~\cite{Ciuchini:2022wbq}, and normalise by the constant $N$
introduced in (\ref{c9mu}) to a new physics scale of 30 TeV. In our analysis of new physics that couples to di-muon pairs, the relevant SMEFT
coefficients are denoted $C_{qe}^{(l) 2322}$ and $C_{lq}^{(l) 2223}$, which are input to
\smelli{} in units of GeV$^{-2}$. We wish to match the SMEFT operators to
include those in (\ref{c9mu}), i.e.
\begin{equation}  \label{eq:cqec9}
	C_{qe}^{(l) 2322} =  N (C_9^{(\mu)} + C_{10}^{(\mu)}), 
\end{equation}
\begin{equation}	
	C_{lq}^{(l) 2223} =  N (C_9^{(\mu)} - C_{10}^{(\mu)}). \label{eq:cqec10}
\end{equation}
These are WCs multiplying the dimension-6 SMEFT operators in the Lagrangian density: 
\begin{equation}
	O_{qe}^{(l) 2322} = (\bar{Q}_2 \gamma_{\alpha} Q_3)(\bar{e}_2\gamma^{\alpha}e_2), \\
\end{equation}
	\begin{equation} \label{eq:olq}
	O_{lq}^{(l) 2223} = (\bar{L}_2\gamma_{\alpha}L_2)(\bar{Q}_2\gamma^{\alpha}Q_3), 
\end{equation}
where $L_i$ and $Q_i$ are $SU(2)_L$ doublets and $e_i$ are $SU(2)_L$ singlets. We adapt
Eqs.~\ref{eq:cqec9} -~\ref{eq:olq} to write the equivalent SMEFT coefficients
and operators contributing to the transitions $b\rightarrow se^+e^-$, by
replacing indices $2\rightarrow 1$ on lepton indices and $\mu\rightarrow e$ on weak effective theory coefficients on the right hand sides, in which case the left hand sides read $C_{qe}^{(l) 2311}$, $C_{lq}^{(l) 1123}$, $O_{qe}^{(l) 2311}$ and $O_{lq}^{(l) 1123}$, respectively.

%



We examine constraints from two main categories of observables contained in
\smelli, labelled `quarks' and `LFU'. 
The LFU category consists of 23 measurements from Belle, LHCb and BaBar which
include constraints from the ratios $R_K$ and $R_{K^*}$ (including the
updates by 
LHCb in 2022). We aim to understand to which extent the updated measurements favour adding
new physics couplings to di-electron pairs. Therefore, we additionally examine
these two LFU ratios separately from the total set of LFU contributions in our
results. The `quarks' category contains 224 other contributions from LHCb
measurements of $B$ meson decays and other similar measurements from ATLAS,
CMS, Belle and BaBar. 

The \smelli\ package requires several tools for performing a
phenomenological analysis, including \flavio\ for computing flavour and
other precision observables and accounting for their theory uncertainties,
alongside \wilson\ for matching between the weak effective theory and the
SMEFT and performing the renormalisation group running. The combination of
these and other tools allows \smelli\ to produce a SMEFT likelihood function
including a total of 247 observables to compare with
predictions~\cite{Stangl:2020lbh}.

Our global fits aim to identify the preferred ranges of 
WCs parameterising new physics by performing a $\chi^2$ test
(as described in Ref.~\cite{Barlow:2019svl}). Combined measurements include
relevant sectors of experimental physics as
including $B$-decay
and LFU violating
observables.
By using \smelli\ for our predictions, we take into account
the mixing between different sectors under renormalisation.

A similar fit to one of the four that we present in this section has also been
performed (with a somewhat
different set of $b-$observables and a different calculation of the 
predictions of observables) in Ref.~\cite{Alguero:2023jeh}\footnote{The fit of
Ref.~\cite{Alguero:2023jeh} goes beyond QCD factorisation, allowing it to
use the $q^2 \in [6,8]$ bin of various measurements, unlike our fit
using \flavio{}, which uses QCD factorisation for its predictions and so
excludes that bin.}. Here, we present our results as a 
function of $C_{9,10}^{(e,\mu)}$ for ease of comparison, even though actually
our fit involves additional and related operators (implied by
SMEFT) that are related by $SU(2)_L$ symmetry. 

\subsection{Fit results} \label{sec:fitresults}

\begin{figure}
	\begin{center}
	  \includegraphics[width=0.45 \textwidth]{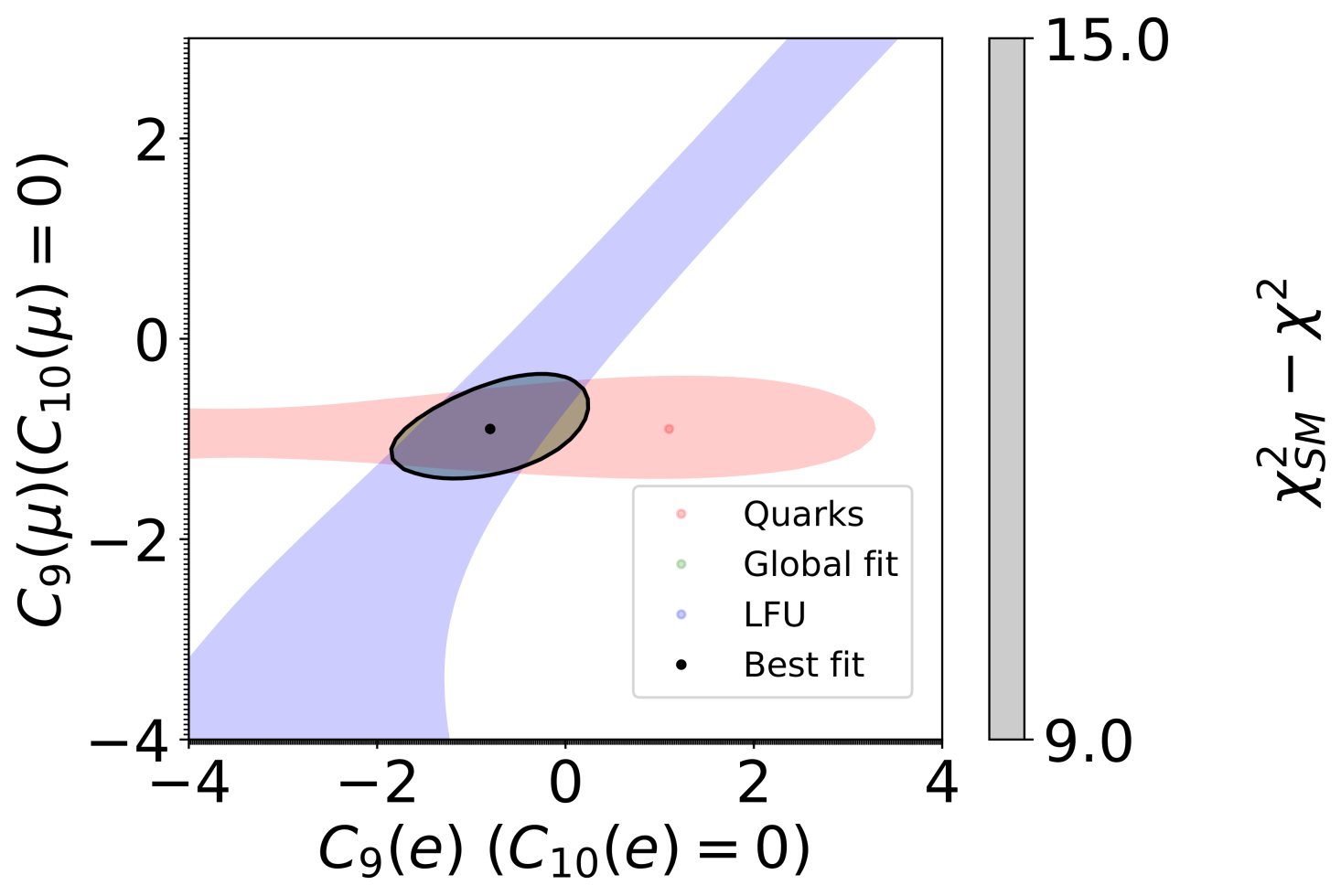}
          \includegraphics[width=0.45 \textwidth]{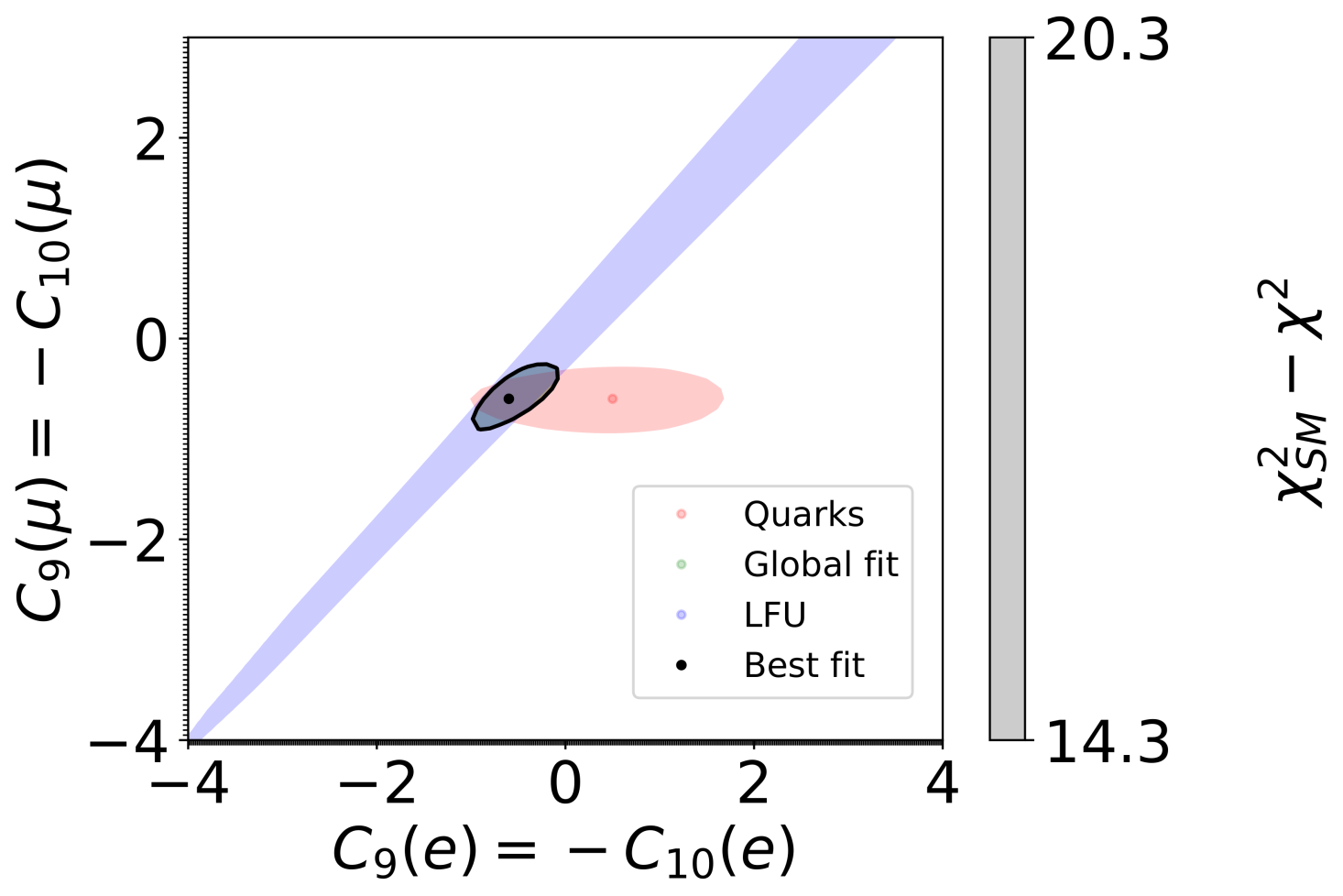} \\
	  \includegraphics[width=0.45 \textwidth]{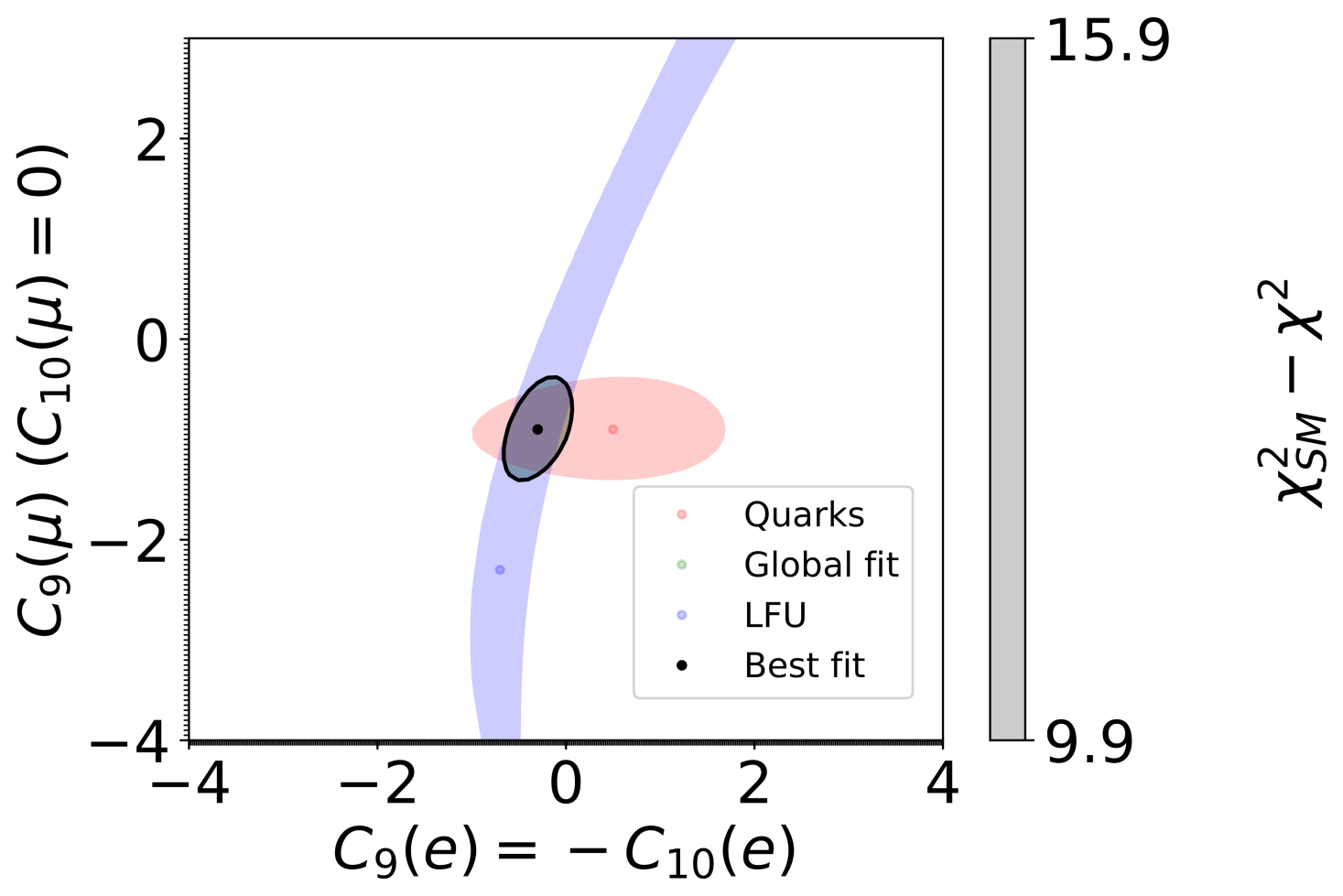}
          \includegraphics[width=0.45 \textwidth]{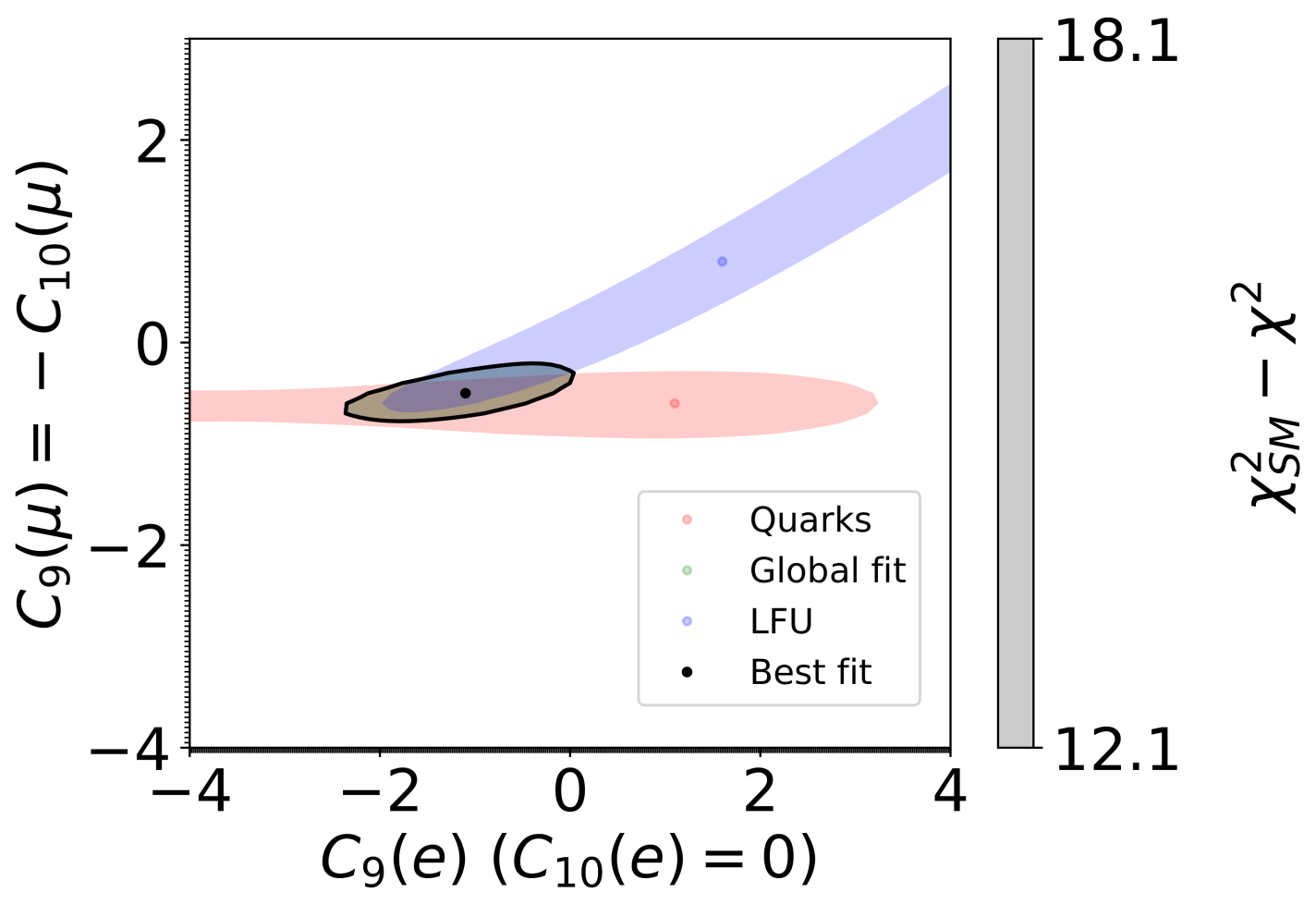}
	\end{center}
	\caption{95$\%$ confidence level (CL) fit regions in the SMEFT parameter
          space where 
          (top-left) $C_{10}^{(e)} = C_{10}^{(\mu)} = 0$,
          (top-right) $C_9^{(\mu)} = -C_{10}^{(\mu)}$ and $C_9^{(e)} =
          -C_{10}^{(e)}$,
          (bottom-left) $C_9^{(e)} = -C_{10}^{(e)}$ and $C_{10}^{(\mu)}=0$,
          (bottom-right) $C_{10}^{(e)}=0$ and $C_9^{(\mu)} = -C_{10}^{(\mu)}$
          via (\ref{eq:cqec9}), (\ref{eq:cqec10}).
          `LFU' contains 23 measurements from Belle, LHCb and BaBar which
          test lepton flavour universality. `quarks'
          contains the 224 other \bsmm\ measurements defined in \flavioA. The
          grey colour bar on the right-hand side is labelled with the extremes
          of the $\chi^2_{SM}-\chi^2$ values within the 95$\%$ `global'
          contour. A positive value of $\chi^2_{SM} -
          \chi^2$ indicates an \emph{improvement} of the fit with respect
          to the SM\@. 
          The best fit is shown by the black point.
  	      \label{fig:c9eVSc9mu_case1}}  
\end{figure}

In our first scenario, we set $C_{10}^{(e)} = C_{10}^{(\mu)} = 0$ and allow
$C_9^{(e)}$ and $C_9^{(\mu)}$ to vary freely, corresponding to the case where
new physics has vector-like couplings to both di-muon and di-electron
pairs. 
The result is plotted in Fig.~\ref{fig:c9eVSc9mu_case1} (top-left). A significant
region of overlap exists between the `LFU' and `quarks' constraints where
$C_9^{(e)}$ takes values between around -2 and its SM value of 0. The most
constraining observables from the collection of 23 that test lepton flavour
universality appear to be $R_K$ and $R_{K^\star}$. This fit was performed first
by Ref.~\cite{Alguero:2023jeh} (using a different
theoretical calculation of the SM prediction and theoretical uncertainties)
and our \flavioA\ fit shows a rather similar 95$\%$ CL region of global
fit\footnote{The $\chi^2$ improvement upon the SM is significantly higher in
Ref.~\cite{Alguero:2023jeh} due mainly to the inclusion there of the $q^2 \in
[6,8]$ GeV$^2$ bin.}. 

The second scenario we consider here requires $C_9^{(e)} = -C_{10}^{(e)}$ and
$C_9^{(\mu )} = -C_{10}^{(\mu )}$ such that di-electron and di-muon pairs have
only left-handed couplings with new physics, leaving a smaller range of
best-fit values for $C_9^{(e)}$ as shown in Fig.~\ref{fig:c9eVSc9mu_case1}
(top-right).   

Another possibility we consider is that of vector-like couplings to
di-muon pairs and left-handed couplings of new physics to di-electron pairs, presented in Fig.~\ref{fig:c9eVSc9mu_case1} (bottom-left). The range of best-fit values for
$C_9^{(e)}$ 
is similar here to that in the top-right panel, but this
scenario includes a wider range of best-fit values for $C_9^{(\mu)}$.  

The final scenario we consider is Fig.~\ref{fig:c9eVSc9mu_case1}
(bottom-right) where the couplings to new physics are swapped compared with
the scenario in the bottom-left panel such that di-muon pairs have left-handed
couplings and di-electron pairs have vector-like couplings. A larger range of
good-fit values for $C_9^{(e)}$ result for this scenario; it is a case with a
fit that includes values extending below $C_9^{(e)}=-1$. 

The main outcome of Ref.~\cite{Alguero:2023jeh} is to evaluate global fits
with and without new physics contributions to electron modes under a framework
containing updates to both experimental measurements and theoretical
calculations of form factors. Within this fully updated framework, the
results in that reference identify that new physics introduced by
$C_9^{(\mu)}$ is mildly preferred over 
scenarios with $C_9^{(\mu)} = -C_{10}^{(\mu)}$, favouring
a vector-like coupling to  di-muon pairs over a left-handed coupling. The fit
also reveals that data can be compatible with non-zero $C_{10}^{(\mu)}$,
although support for these scenarios is not as strong.

Other analyses support the introduction of non-zero new physics WCs, though
with different assumptions to ours. For example, a recent evaluation 
including possible new physics WCs~\cite{Wen:2023pfq}
performed higher dimensional global fits for several more non-zero WCs
instead of focusing on the four that we examine here.
There is therefore no overlap between our results and those of
Ref.~\cite{Wen:2023pfq}. Another fit assumes that new physics affects electrons
and muons identically~\cite{SinghChundawat:2022zdf}, an assumption which we do
not follow in the present paper. 

Both Refs.~\cite{Alguero:2023jeh} and~\cite{Wen:2023pfq} provide insight into
a renewed focus on LFU new physics by examining the differences between
global fits before and after the release of the 2022 LHCb update of $R_{K}$
and $R_{K^{\star}}$. The large impact of such observables on constraining
the parameter plane $C_9^{(e)} - C_9^{(\mu)}$ can be seen in 
Fig.~\ref{fig:c9eVSc9mu_case1}.

Fig.~\ref{fig:c9eVSc9mu_case1} indicates that the a best-fit point has
$C_9^{(e)}\approx C_9^{(\mu)}\neq 0$,
but that $C_9^{(e)}=0$ can also fit the data, as shown in the top-left hand
panel. Motivated by this top-left hand panel and similar previous results in Ref.~\cite{Alguero:2023jeh} we shall now turn to a set of 
models which interpolates between $C_9^{(e)}=C_9^{(\mu)}\neq 0$ and
$C_9^{(e)}=0$ (and which also extrapolates outside of these constraints). 


\section{Models \label{sec:models}}

\begin{figure}
  \begin{center}
    \includegraphics[width=85pt]{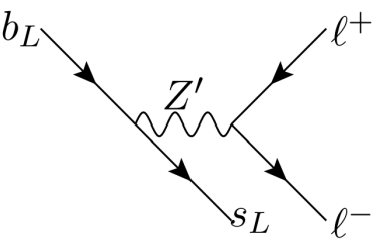}
  \end{center}
  \caption{\label{fig:NCBAs} Feynman diagram of the leading new
    physics 
    contribution to \bsll\ observables from a suitable $Z^\prime$ model.}    
\end{figure}

Our $U(1)_X$ gauge symmetry (which is extra to the gauge symmetry of the SM)
is expected to be spontaneously broken  
by a complex scalar `flavon' field $\theta$, whose $U(1)_X$ charge $Q$ is
non-zero.
Of the models we shall propose, aspects such as these just mentioned are very
similar to the $U(1)_{Y_3-x(L_\mu-L_\tau)}$ model~\cite{Davighi:2021oel}, the
$U(1)_{B_3-L_2}$ model~\cite{Bonilla:2017lsq,Alonso:2017uky,Chun:2018ibr,Allanach:2020kss}, Third Family Hypercharge
models (TFHMs)~\cite{Allanach:2018lvl,Allanach:2019iiy},  or
mixtures between the $U(1)_{B_3-L_2}$ model and the TFHM~\cite{Allanach:2022bik}.
The massive electrically neutral gauge boson resulting from Higgsed 
$U(1)_X$ breaking is dubbed a
$Z^\prime$ boson which has a mass
\begin{equation}
  M_{Z^\prime} = Q g_{Z^\prime} \langle \theta \rangle,
\end{equation}
where $\langle \theta \rangle$ is the vacuum expectation value of the flavon
field. Whichever fields possess non-zero $U(1)_X$ charges will generically
have couplings to the $Z^\prime$ boson. Thus we wish second-family leptons to
have a non-zero charge,
and (following the 
arguments in \S\ref{sec:intro}), 
possibly first-family leptons as well\footnote{Even with the $Z^\prime$ \emph{not}
coupling to electrons, the fit of flavour data can be significantly improved
with respect to that of the SM~\cite{Allanach:2022iod,Fleischer:2023zeo}.}.
We also wish the third family of quarks to
have a charge in order to establish a $Z^\prime$ coupling to $s \bar b +
H.c.$, starting from a coupling to $b \bar b$ in the weak eigenbasis, as also
explained in \S\ref{sec:intro}. Then, we may explain the \bsll\ anomalies by
new physics contributions to the amplitude
like the one depicted in Fig.~\ref{fig:NCBAs}.
It should be evident from the above discussion that the charges of the various
fields in the model affect the phenomenology of it, since they determine what
the $Z^\prime$ couples to (and with which relative strength). Therefore, we now
discuss the fermionic charge assignment of the model. 

\subsection{Fermionic charge assignment}
We begin by extending the SM by three right-handed neutrino fields
$\nu_i$. These $\nu_i$ fields will be used both for anomaly cancellation, 
with an eye to ultimately providing an ultra-violet model of neutrino masses
(this we 
shall not specify in any detail, however). The chiral fermionic field content
of the model and its representations under the SM$\times U(1)_X$ gauge group
is specified in Table~\ref{tab:charges}. We note here that for brevity we shall denote both the field and its $U(1)_X$ charge by the same label;
context should make the meaning of the symbol clear.
\begin{table}
  \begin{center} 
    \begin{tabular}{|c|c|} \hline
        Field & $(SU(3), SU(2), U(1)_Y, U(1)_X)$ \\ \hline
        $Q_i$  & $(3, 2, 1, Q_i)$ \\
        $L_i$  & $(1, 2, 3, L_i)$ \\
        $u_i$  & $(3, 1, 4, u_i)$ \\
        $d_i$  & $(3, 1, -2, d_i)$ \\
        $e_i$  & $(1, 1, -6, e_i)$ \\
        $\nu_i$& $(1, 1, 0, \nu_i)$ \\
        \hline
    \end{tabular}
    \caption{\label{tab:charges} Representations of fermionic chiral fields
      under the SM$\times U(1)_X$ gauge group. $i \in \{1,2,3\}$ is a family
      index and gauge indices have been suppressed. $Q_i$ and $L_i$ are
      left-handed Weyl fermions, whereas the other fields listed are all
      right-handed Weyl fermions. We have chosen here to normalise the
      hypercharge gauge coupling so that the hypercharges of all fermionic
      fields are integers.} 
    \end{center}
  \end{table}

As explained above, we want to find a set of models that can potentially address
the \bsmm\ anomalies, but which interpolate between models where the
$Z^\prime$ does \emph{not} couple to electrons and those where it couples
with an equal strength to di-electron pairs and di-muon pairs.
However, the $U(1)_X$ charge assignments
have constraints upon them given by the requirement of not generating quantum
field theoretic anomalies, which would spoil the gauge symmetry.
We shall now
go through the arguments that lead us to the charge assignments, since they
shall make clear to which extent the assignments are constrained by anomaly
cancellation, to which extent they are a choice dictated by desired
phenomenology and the extent to which they are just a choice to be concrete.

We first list the anomaly cancellation conditions which a gauged $U(1)_X$
chiral-fermion charge assignment should respect~\cite{Allanach:2018vjg}:
\begin{eqnarray}
  \sum_i \left(2  Q_i - u_i - d_i\right) &=& 0, \label{X33}\\
  \sum_i \left(L_i + 3 Q_i \right)      &=& 0, \label{X22}\\
  \sum_i \left( Q_i + 3L_i - 8 u_i - 2 d_i - 6 e_i\right) &=& 0, \label{XYY}\\
  \sum_i \left( 6 Q_i + 2 L_i - 3 u_i - 3 d_i - e_i - \nu_i \right) &=&
  0, \label{Xgrav} \\
  \sum_i \left( Q_i^2 - L_i^2 - 2 u_i^2 + d_i^2 + e_i^2\right) &=& 
  0    \label{quad}  \\
  \sum_i \left( 6 Q_i^3 +2 L_i^3 - 3 u_i^3 -3 d_i^3 - e_i^3 - \nu_i^3\right) &=& 0.   \label{cub}  
\end{eqnarray}
These equations have been simultaneously solved over the integers
both numerically with $U(1)_X$ charges
between 10 and -10~\cite{Allanach:2018vjg}  and, more generally,
analytically~\cite{Allanach:2020zna}. Rather than begin with these solutions
and then restrict them, 
we instead find it instructive to make
some choices based partly on expected phenomenological consequences of some
charge assignments whilst simultaneously applying (\ref{X33})-(\ref{cub}).

We require a coupling of the $Z^\prime$ to left-handed $b \bar s$
quark pairs in order to explain an apparent new physics effect in
\bsmm\ transitions. We therefore pick $Q_3 \neq 0$, providing  a $Z^\prime$
coupling to 
left-handed $b \bar b$ pairs and assume that its coupling to (left-handed)
$\bar b
s + H.c.$ will
be provided by some small amount of $b-s$ mixing. The mixing is banned by
$U(1)_X$ but since this is spontaneously broken, we anticipate 
small $U(1)_X$ breaking effects, such as small quark mixing.
We may fix $Q_3=1$ by rescaling the $U(1)_X$ gauge coupling.
If we couple the $Z^\prime$ dominantly 
to the \emph{third} family quarks only, 
direct $Z^\prime$ search bounds from the LHC will be weaker,
since LHC $Z^\prime$ production dominantly occurs via the
$b \bar b \rightarrow Z^\prime$ process,
which is doubly suppressed by both the $b$ and $\bar b$
parton distribution functions. Motivated by this, we fix the $U(1)_X$ charges of
the first two generations of quark fields to zero,
i.e.\ $u_1=u_2=d_1=d_2=Q_1=Q_2=0$. 
Substituting these assignments into 
(\ref{X33}), we obtain that $u_3 + d_3 = 2$. 
We shall here pick $u_3=d_3=1$, meaning that we can characterise the quark
charges in terms of third-family baryon number. (\ref{X22}) then gives that
\begin{equation}
  \sum_i L_i = -3. \label{accQL}
\end{equation}
We shall pick $L_2 \neq 0$ in order to couple the $Z^\prime$
to left-handed muon pairs, since we know from fits to
\bsmm\ anomalies~\cite{Alguero:2023jeh,Wen:2023pfq}
that a new physics contribution to $C_9^{(\mu)}\neq 0$ is necessary to describe 
the pertinent measurements well. 
We shall vary $L_1$ in order to vary the
$Z^\prime$ coupling to left-handed electron pairs: (\ref{accQL}) then can be
rearranged to yield $L_3$.
Substituting (\ref{accQL}) and the other assigned charges into (\ref{XYY}), 
we obtain
\begin{equation}
  \sum_i e_i = -3, \label{eX}
  \end{equation}
which allows us to obtain, using (\ref{Xgrav}), 
\begin{equation}
  \sum_i \nu_i = -3. \label{accNu}
\end{equation}
(\ref{quad}) and (\ref{cub}) then become
\begin{eqnarray}
  \sum_i \left( L_i^2 - e_i^2 \right) = 0, \label{oneacc} \\
  \sum_i \left( 2 L_i^3 - e_i^3 - \nu_i^3\right) = 0. \label{twoacc}
  \end{eqnarray}
(\ref{eX})-(\ref{twoacc}) are solved by 
\begin{equation}
-X_i:=L_i=e_i=\nu_i\ \text{for each}\ i. \label{stip}
\end{equation}
This
is not a general solution of the equations, but it is sufficient for
our purposes here. 
After the stipulation in (\ref{stip}),
there remains only one independent
constraint, which we can take to be (\ref{accQL}). 
(\ref{stip}) allows us to 
summarise the $U(1)_X$ charges in
terms of electron number $L_e$, muon number $L_\mu$ and tau number $L_\tau$.
We shall fix the $X$ charge of $L_2=e_2=\nu_2$ (here dubbed to be $-X_\mu$) to be
a reasonably large integer to allow more resolution in the other charges; we pick $X_\mu=10$. We then allow the $U(1)_X$
electron charge ($-X_e$) to vary. 
The $U(1)_X$ charges of the fermions as a whole can be characterised by
\begin{equation}
  3B_3 - \left(X_e L_e + X_\mu L_\mu + [3 - X_e - X_\mu]L_\tau\right).
  \label{b3mL}
\end{equation}
$X_e/X_\mu=1$ corresponds to the
case where the 
coupling of the $Z^\prime$ to di-electron pairs
is equal to that of di-muon pairs, whereas $X_e=0$ is the case where the
electron does not directly couple to the $Z^\prime$ at tree-level.

The arguments on anomaly cancellation
thus far apply to our assumed chiral fermionic content of the SM
plus three right-handed neutrinos. If one were add a pair of
chiral fermions which are vector-like under the SM gauge symmetries but have
non-cancelling
$U(1)_X$ charges, the system of anomaly equations would change and one could
acquire different solutions to the ones that we have found. One would need to
explain how these 
additional chiral fermions acquire masses to make them significantly heavier
than is probed by current experiments; this might be possible, depending upon
the new chiral fermion charges, by utilising
$U(1)_X$ breaking effects via $\langle \theta \rangle \approx {\mathcal
  O}$(TeV). We note this
caveat, but shall for now assume no additional chiral fermionic
fields of this type. Our anomaly cancellation analysis applies to the chiral
fermionic 
field content in Table~\ref{tab:charges} along with any additional fermions
only being added in vector-like pairs
under the entire SM$\times U(1)_X$ gauge group.

We fix the $U(1)_X$ charge of the SM Higgs doublet $H$ 
so that the top 
Yukawa coupling Lagrangian density term, $\lambda_t \overline{Q_3} H u_3 +
H.c.$, is 
allowed by the gauge symmetry\footnote{Some of the other Yukawa couplings may be disallowed by
$U(1)_X$, but may receive small contributions from
non-renormalisable operators (for example as in the Froggatt-Nielsen
mechanism~\cite{FROGGATT1979277}) once $U(1)_X$ is spontaneously broken.}.
This constraint requires that $H$ then has $U(1)_X$ charge equal to zero,
simplifying our analysis because there is no predicted $Z-Z^\prime$ mixing at
tree-level. Such a mixing would change the predictions of electroweak
precision observables (EWPOs); with zero mixing, as predicted here, we
effectively 
decouple the EWPOs from our discussion.
This is essentially dictated by model choice: in other models, e.g.\ the
TFHMs, the electroweak observables
significantly change with model parameters (the
quality of the electroweak fit in the TFHMs is similar to that of the SM, with
improvements in 
$M_W$ being offset against other EWPOs such as measurements of $Z^0$ boson
couplings to different families of di-lepton
pair~\cite{Allanach:2021kzj}). Thus,  
decoupling the EWPOs as we do here simplifies our analysis but is not
necessarily 
essential phenomenologically: the preference (or otherwise) of electroweak fits
has to be determined on a case-by-case basis.

It behoves us now to specify the other pertinent TeV-scale properties of our
model, which we shall do in the following subsection.

\subsection{More model details}
Here, we deal with the $Z^\prime$-specific parts of the model, which encapsulates
the phenomenology that we are interested in predicting. We shall not find it
necessary to specify all details of the model (the flavon potential or
flavon/Higgs mixing -- for that, see Ref.~\cite{Allanach:2022blr} --
or the
origin of \emph{non-zero} small Yukawa couplings, for example).
The model set-up in the present subsection closely follows that of
Refs.~\cite{Allanach:2018lvl,Allanach:2019iiy,Allanach:2020kss,Allanach:2022bik},
which are discriminated from the present model by the fermionic
$U(1)_X$ charge assignments.  
The model is
supposed to be at the level of a TeV-scale effective field theory that
includes the quantum fields of the SM, three right-handed neutrino fields and
the $Z^\prime$. We write the fermionic fields in the gauge eigenbasis with a
primed notation
\begin{eqnarray}
{\bf u_L'}&=&\left( \begin{array}{c} u_L' \\ c_L' \\ t_L' \\ \end{array}
\right), \qquad
{\bf d_L'}=\left( \begin{array}{c} d_L' \\ s_L' \\ b_L' \\ \end{array}
\right), \qquad
{\bf e_L'}=\left( \begin{array}{c} e_L' \\ \mu_L' \\ \tau_L' \\ \end{array}
\right), \qquad
{\bm \nu_L'}=\left( \begin{array}{c} {\nu_e'}_L \\ {\nu_\mu'}_L
  \\ {\nu_\tau'}_L \\ \end{array} \right),
 \nonumber \\ 
{\bf u_R'}&=&\left( \begin{array}{c} u_R' \\ c_R' \\ t_R' \\ \end{array}
\right), \qquad
{\bf d_R'}=\left( \begin{array}{c} d_R' \\ s_R' \\ b_R' \\ \end{array}
\right),\qquad
{\bf e_R'}=\left( \begin{array}{c} e_R' \\ \mu_R' \\ \tau_R' \\ \end{array}
\right), \qquad
{\bm \nu_R'}=\left( \begin{array}{c} {\nu_e'}_R \\ {\nu_\mu'}_R
  \\ {\nu_\tau'}_R \\ \end{array} \right),
\end{eqnarray}
along with the SM fermionic electroweak doublets
\begin{equation}
{\bf Q'}_i=\left( \begin{array}{c} {\bf u_L'}_i \\ {\bf d_L'}_i \end{array}
\right),\qquad
{\bf L'}_i=\left( \begin{array}{c} {\bm \nu_L'}_i \\ {\bf e_L'}_i \end{array}
\right).  
\end{equation}
The neutrinos and SM fermions acquire masses after the SM Brout-Englert-Higgs mechanism through 
\begin{eqnarray}
-\mathcal{L}_{Y}&=&\overline{\bf Q'}  Y_u \tilde H {\bf u'_R} +
\overline{\bf Q'}  Y_d H  {\bf d'_R} +
\overline{\bf L'}  Y_e H  {\bf e'_R} + 
\overline{\bf L} Y_\nu \tilde H {\bm \nu'_R} + 
\frac{1}{2}{\overline{{\bm \nu_R'}^c}} M 
  {\bm \nu_R'} + H.c. , 
 \label{yuk}
\end{eqnarray}
where $Y_u$, $Y_d$ and $Y_e$ are dimensionless complex coupling constants,
each written as a 3 by 3 matrix in family space; the matrix $M$ is a 3 by 3
complex symmetric matrix of mass dimension 
1, ${\Phi}^c$ denotes the
charge conjugate of field $\Phi$ and $\tilde H := ({H^0}^\ast, -H^-)^T$. 
Gauge indices have been
omitted in (\ref{yuk}).

For $X_e \neq X_\mu \neq X_\tau$, the Yukawa couplings have the following
textures at the renormalisable tree level in the unbroken $U(1)_X$ limit:
\begin{eqnarray}
  Y_u \sim \begin{pmatrix}
    \times & \times & 0 \\
    \times & \times & 0 \\
    0      &      0 & \times \\
  \end{pmatrix}, \quad
  Y_d \sim \begin{pmatrix}
    \times & \times & 0 \\
    \times & \times & 0 \\
    0      &      0 & \times \\
  \end{pmatrix}, \quad
  Y_e \sim \begin{pmatrix}
    \times & 0 & 0 \\
    0 & \times & 0 \\
    0      &      0 & \times \\
  \end{pmatrix}. \label{textures}
\end{eqnarray}
The structure of $Y_e$ predicts that charged-lepton violating
couplings of the $Z^\prime$ are zero. 

We may write
$H=(0,\ (v + h)/\sqrt{2})^T$ after electroweak symmetry breaking,
where 
$h$ is the physical Higgs
boson field
and (\ref{yuk}) includes the fermion mass terms 
\begin{eqnarray}
-\mathcal{L}_{Y}&=&\overline{\bf u'_L} V_{u_L} V_{u_L}^\dagger m_u V_{u_R}
V_{u_R}^\dagger {\bf u'_R} + 
\overline{\bf d'_L} V_{d_L} V_{d_L}^\dagger m_d  V_{d_R} 
V_{d_R}^\dagger {\bf d'_R} + 
\overline{\bf e'_L} V_{e_L} V_{e_L}^\dagger m_e  V_{e_R} 
V_{e_R}^\dagger {\bf e'_R} + \nonumber \\ &&
\frac{1}{2} ( {\overline{\bm \nu_L'}}\ \overline{{\bm \nu_R'}^c}) M_\nu
\left( \begin{array}{c} {\bm {\nu_L'}}^c \\ {\bm \nu_R'} \\
\end{array}
  \right)
  +H.c. + \ldots, \label{diracMass}
\end{eqnarray}
where
\begin{equation}
M_\nu = \left( \begin{array}{cc} 0 & m_{\nu_D} \\
  m_{\nu_D}^T & M \\ \end{array} \right),
\end{equation}
$V_{I_L}$ and $V_{I_R}$ are 3 by 3 unitary mixing matrices for each
field species $I$, 
$m_u:=v Y_u/\sqrt{2}$, $m_d:=v
Y_d/\sqrt{2}$, $m_e:=v Y_e/\sqrt{2}$ and $m_{\nu_D}:=v Y_\nu/\sqrt{2}$, where
$v$ is the SM Higgs expectation value, measured to be 246.22
GeV~\cite{ParticleDataGroup:2022pth}. 
The final explicit term in (\ref{diracMass})
incorporates the see-saw
mechanism via a 6 by 6 complex
symmetric mass matrix. Since the elements in $m_{\nu_D}$ are much smaller than
those in $M$, we perform a rotation to obtain a 3 by 3 complex symmetric
mass matrix for the three light neutrinos. These approximately
coincide with the left-handed weak eigenstates ${\bm \nu'_L}$, whereas
three heavy neutrinos approximately correspond to the right-handed weak
eigenstates ${\bm \nu'_R}$. The neutrino mass term of (\ref{diracMass}) becomes, to a good
approximation, 
\begin{equation}
-  {\mathcal L}_{\nu} =
  \frac{1}{2} {\overline {\bm \nu_L'^c}} m_\nu {\bm \nu_L'} +
\frac{1}{2} {\overline {\bm \nu_R'^c}} M {\bm \nu_R'} + H.c., 
  \end{equation}
where $m_\nu:= m_{\nu_D}^T M^{-1} m_{\nu_D}$ is a complex symmetric 3 by 3
matrix. 

Choosing
$V_{I_L}^\dagger m_I  V_{I_R}$ to be diagonal, real and positive for $I
\in \{ u,d,e\}$, and
$V_{{\nu}_L}^T m_\nu  V_{{\nu}_L}$ to be diagonal,
real and positive 
(all in ascending order of mass
from the top left toward the bottom right of the matrix), we can identify the 
{\em non}-primed {\em mass}\/ eigenstates\footnote{${\bf P}$ and ${\bf P}'$ are column 3-vectors.}
\begin{equation}
  {{\bf P}} = V_P^\dag {\bf P'} \ \text{where}\ {P} \in \{{u_R},\
  {d_L},\ {u_L},\ {e_R},\ {u_R},\ {d_R},\ {\nu}_{L},\ {e_L}\}. \label{fermion_rotations}  
  \end{equation}
We may then find the CKM matrix $V$ and the
Pontecorvo-Maki-Nakagawa-Sakata (PMNS) matrix $U$ in terms of the fermionic
mixing matrices:
\begin{equation}
V=V_{u_L}^\dagger V_{d_L}, \qquad U = V_{\nu_L}^\dagger V_{e_L}. \label{mix}
\end{equation}
The zeroes in $Y_u$ and $Y_d$ in (\ref{textures}) predict that the magnitudes
of the elements 
$V_{23}, V_{13}, V_{32}, V_{31}$ are much smaller than 1, agreeing with
measurements~\cite{ParticleDataGroup:2022pth}. Clearly, more
model building into the 
ultra-violet would be required to understand the details of neutrino masses
and mixing and how exactly the zeroes in (\ref{textures}) are filled in
with 
small entries. We leave such model considerations aside, instead pointing to
\S{}2.5 of Ref.~\cite{Allanach:2022iod} for some possibilities.

The kinetic terms of the $U(1)_X$ gauge boson yield the following 
$Z^\prime$ interactions
\begin{equation}
  {\mathcal L}_I = - g_{Z^\prime} \left(\overline{\bf Q'} \slashed{Z}^\prime \xi {\bf Q'} +
  \overline{\bf u'_R} \slashed{Z}^\prime \xi {\bf u'_R} + \overline{d'_R}
  \slashed{Z}^\prime \xi {\bf d'_R} +
  \overline{\bf L'} \slashed{Z}^\prime \Xi {\bf L'} + \overline{\bf e'_R}
  \slashed{Z}^\prime \Xi {\bf e'_R} +
  \overline{{\bm \nu'_R}} \slashed{Z}^\prime \Xi {{\bm \nu'_R}}
  \right), \label{ints}
\end{equation}
where
\begin{equation}
  \xi := \begin{pmatrix}
    0 & 0 & 0 \\
    0 & 0 & 0 \\
    0 & 0 & 1 \\
  \end{pmatrix}, \qquad
  \Xi := \begin{pmatrix}
    -X_e & 0 & 0 \\
    0 & -X_\mu & 0 \\
    0 & 0 & -X_\tau \\
  \end{pmatrix}
\end{equation}
are fixed by the fermionic fields' $U(1)_X$ charges.
In the unprimed mass eigenbasis, (\ref{ints}) becomes
\begin{eqnarray}
{\mathcal L} _I&=& - g_{Z^\prime} \left(\overline{\bf u_L} \slashed{Z}^\prime \Lambda_\xi^{u_L}{\bf
  u_L} +
\overline{\bf d_L} \slashed{Z}^\prime\Lambda_\xi^{d_L} {\bf d_L} +
  \overline{\bf u_R} \slashed{Z}^\prime\Lambda_\xi^{u_R} {\bf u_R} +
  \overline{\bf d_R} \slashed{Z}^\prime\Lambda_\xi^{d_R} {\bf d_R} +
  \right. \nonumber \\ && \left.
  \overline{\bf e_L} \slashed{Z}^\prime\Lambda_\Xi^{e_L} {\bf e_L} +
  \overline{\bf e_R} \slashed{Z}^\prime\Lambda_\Xi^{e_R} {\bf e_R} +
  \overline{\bm \nu_L} \slashed{Z}^\prime\Lambda_\Xi^{\nu_L} {\bm \nu_L} +
  \overline{\bm \nu_R} \slashed{Z}^\prime\Lambda_\Xi^{\nu_R} {\bm \nu_R}
  \right), \label{intsMass}  
\end{eqnarray}
where $\Lambda_{\alpha}^{P}:=V_{P}^\dag \alpha V_{P}$ for 
$\alpha \in \{ \xi, \Xi \}$. 
The right-handed neutrinos ${\bm \nu'_R}$ are assumed to be heavy compared to
the TeV scale and play no further role in the phenomenology of
\bsmm\ anomalies; we shall therefore neglect them in the discussion that
follows. 

To make phenomenological progress with our models, we shall need to specify
$V_{P}$. We simply assume that the ultra-violet model details are such
that the zeroes in (\ref{textures}) are filled in (or not) at the correct
level for experiment. 
The $V_P$ are 3 by 3
unitary matrices, and we
pick a simple ansatz which is not immediately obviously ruled out by 
strong flavour changing neutral current constraints on charged lepton flavour
violation or neutral current flavour violation in the first two families of
quark. Firstly, we set $V_{e_R}=V_{d_R}=V_{u_R}=V_{e_L}=I$, the 3 by 3 identity
matrix.
A non-zero $(V_{d_L})_{23}$ matrix element is required for  the $Z^\prime$ to
mediate new physics contributions to $b\rightarrow s \ell^+\ell^-$ transitions.
We capture the important quark mixing (i.e.\ 
between $s_L$ and $b_L$) in $V_{d_L}$ as
\begin{equation}
  V_{d_L} = \left(\begin{array}{ccc} 1 & 0 & 0 \\
  0 & \cos \theta_{sb} & \sin \theta_{sb} \\
  0 & -\sin \theta_{sb} & \cos \theta_{sb} \\ \end{array}
  \right). \label{ansatz}
\end{equation}
$V_{\nu_L}$ and $V_{u_L}$ are fixed by (\ref{mix}), where we use the
experimentally determined values for the entries of $V$ and $U$ via the
central values in the standard parameterisation from
Ref.~\cite{ParticleDataGroup:2022pth}.
Having fixed all of the fermionic
mixing matrices, we have provided an ansatz that could be perturbed around for
a more complete characterisation. We leave such perturbations
aside in the present paper.

Here, we summarise the SMEFT operators that result from integrating out the
$Z^\prime$; they are given in Table~\ref{tab:smeft_ops}, ready for input into
\flavio{}~\cite{Straub:2018kue}.
\begin{table}
  \begin{center}
    \begin{tabular}{|cc||cc|} \hline
      WC                         & value                        & WC
      & value               \\ \hline
      $C_{ll}^{iiii}$              & $-\frac{1}{2} X_i^2$          & $C_{ll}^{iijj}$ $(i\neq j)$ & $-X_iX_j$            \\
      $(C_{lq}^{(1)})^{iijk}$       & $X_i( \Lambda_\xi^{d_L})_{jk}$ & &  \\
      $C_{ee}^{iijj}$ $(i \neq j)$ & $-X_i X_j$                    &
      $C_{uu}^{3333}$             & $-\frac{1}{2}$       \\
      $C_{dd}^{3333}$             & $-\frac{1}{2}$ & $C_{ee}^{iiii}$             & $-\frac{1}{2}X_i^2$    \\            
      $C_{eu}^{ii33}$              & $X_i$                         & $C_{ed}^{ii33}$              & $X_i$               \\
      ${C_{ud}^{(1)}}^{3333}$       & $-1$                           & $C_{le}^{iijj}$              & $-X_i X_j$           \\
      $C_{qe}^{ijkk}$              & $X_k (\Lambda_\xi^{d_L})_{ij}$        & ${C_{qu}^{(1)}}^{ij33}$       & $-(\Lambda_\xi^{d_L})_{ij}$ \\
      ${C_{qd}^{(1)}}^{ij33}$       & $-(\Lambda_\xi^{d_L})_{ij}$ &
      ${C_{qq}^{(1)}}^{ijkl}$       & $(\Lambda_\xi^{d_L})_{ij}(\Lambda_\xi^{d_L})_{kl} \frac{\delta_{ik}\delta_{jl}-2}{2}$  \\
      $C_{lu}^{ii33}$       & $X_i$                           &
      $C_{ld}^{ii33}$              & $X_i$           \\
      \hline
    \end{tabular}
  \end{center}
  \caption{\label{tab:smeft_ops} Non-zero $M_{Z^\prime}$-scale SMEFT operators
    in units of $g_{Z^\prime}^2 / M_{Z^\prime}^2$, in terms of the left-handed
    lepton doublet $U(1)_X$ charges $-X_i=L_i$ for $i \in \{1,2,3\}$. The
    notation is in the down-aligned 
    Warsaw basis~\cite{Grzadkowski:2010es}. There is no sum implied upon repeated
    family indices $i,j,k,l \in \{1,2,3\}$.} 
  \end{table}
We note that, to specify the model and its $Z^\prime$ phenomenology, once we
have picked a value for $X_e$, there are three important model
parameters that affect the pertinent phenomenology: $g_{Z^\prime}$,
$M_{Z^\prime}$ and $\theta_{sb}$, but at tree-level, as
Table~\ref{tab:smeft_ops} shows, the flavour data only depend upon two
effective parameters: the combination $g_{Z^\prime}/M_{Z^\prime}$ and $\theta_{sb}$. 

\section{LEP constraints \label{sec:LEP}}
\begin{figure}
  \begin{center}
    \includegraphics[width=200 pt]{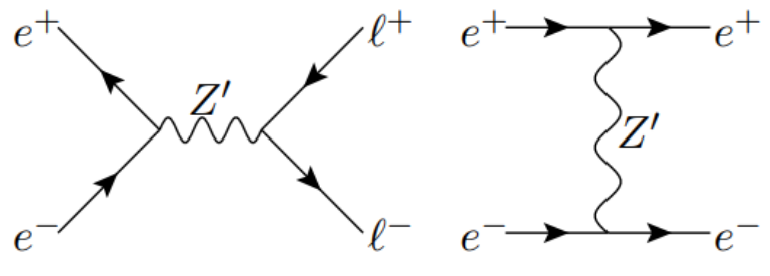}
  \end{center}
  \caption{\label{fig:ee} Feynman diagram of the leading $Z^\prime$
    contribution to LEP-2 di-lepton production from a suitable 
    model. Here, $\ell \in \{ e, \mu, \tau \}$.} 
\end{figure}

Since some of the models we consider couple the $Z^\prime$ to
di-electron pairs, LEP-2 
di-lepton production cross-section measurements, which are broadly in
agreement with SM predictions, provide constraints. 
This is because a contribution from the $Z^\prime$ becomes non-zero: some
leading Feynman diagrams for its contribution to the amplitude are shown in
Fig.~\ref{fig:ee}.
In this section, we shall re-examine the constraints on four-lepton dimension-6
SMEFT WCs coming from LEP-2. Analytic expressions for the expected dominant
contributions from these (in the interference terms) have been already
calculated in Ref.~\cite{Falkowski:2015krw}. Here, we re-calculate the full
dependence of the tree-level predictions upon the WCs ready
for inclusion into \flavio{}. By extracting the interference terms,
we shall provide an independent check upon the analytic results presented in
Ref.~\cite{Falkowski:2015krw}. By calculating the \emph{full} tree-level
dependence upon the WCs
(i.e.\ not only including the interference terms)
and putting them into \flavio{}, we evade possible computational problems with
predicted negative cross-sections when performing parameter
scans. Ref.~\cite{Falkowski:2015krw} provided numerical results of a fit to
electroweak measurements of the epoch
and other LEP measurements, where some of the WCs are constrained
at the $\mathcal{O}(10^{-2})/v^2$ level, where $v=246.22$ GeV is the SM Higgs
vacuum expectation value. Although LEP experimental measurements have not
changed since Ref.~\cite{Falkowski:2015krw}, some electroweak data
have. Providing the LEP constraints as part of the \flavio\ package should
then facilitate SMEFT fits in general as well as 
fits to our $Z^\prime$ models, once we have matched the models to the SMEFT. 

Some of the SMEFT WCs alter differential scattering
cross-section predictions of $e^+ e^- \rightarrow \mu^+ \mu^-$, $e^+ e^-
\rightarrow \tau^+ \tau^-$ and $e^+ e^- \rightarrow e^+ e^-$ (Bhabha
scattering).  
In the Warsaw convention~\cite{Grzadkowski:2010es}, the relevant
WCs which can alter the predictions for these processes 
are $C_{le}^{1jj1}$, $C_{ll}^{11jj}$, $C_{ee}^{11jj}$, $C_{le}^{11jj}$
and $C_{le}^{11jj}$, where $j \in \{1,2,3\}$.
The predictions for $e^+ e^- \rightarrow \mu^+ \mu^-$ and $e^+ e^- \rightarrow
\tau^+ \tau^-$ are simple and almost identical to each other 
and so we consider them first, before going on to consider Bhabha scattering.

\subsection{LEP: di-muon and di-tau final states}
Building notation similar to that in Ref.~\cite{Greljo:2022jac}, 
we consider the tree-level polarised scattering amplitudes of massless
di-electron pairs into either
massless di-muon pairs ($j=2$) or massless di-tau pairs ($j=3$) including
4-lepton dimension-6 SMEFT operators 
\begin{equation}
  {\mathcal M}\left(e^+ e^- \rightarrow e_j^+ e_j^-\right) =
  -i \left(\bar e \gamma^\alpha P_L e_j\right)\left(\bar e_j \gamma_\alpha P_R e
  \right) C_{le}^{1jj1} +
  i \sum_{X,Y}
  \left(\bar e \gamma^\alpha P_X e \right) \left(\bar e_j \gamma_\alpha P_Y
  e_j\right) N_{1jj1}^{XY}(s),
\end{equation}
where the sum is over $\{X,Y\} \in \{L, R\}$,
\begin{eqnarray}
  N_{1jj1}^{XY}(s) &:=&   
  \frac{e^2}{s} + \frac{g_Z^{e_X}g_Z^{{e_j}_Y}}{s - M_Z^2 + i \Gamma_Z M_Z} +
   \left(C_{ll}^{11jj} + C_{ll}^{1jj1}\right) \delta_{XL}\delta_{YL} +
   C_{ee}^{11jj}\delta_{XR}\delta_{YR} + 
   \nonumber  \\ && 
   C_{le}^{11jj} \delta_{XL} \delta_{YR} +
    C_{le}^{jj11} \delta_{XR} \delta_{YL} ,
\end{eqnarray}
$\delta_{LL}=\delta_{RR}=1,\ \delta_{LR}=\delta_{RL}=0$, 
$s,t$ and $u$ are the usual Mandlestam kinematic variables,
$M_Z$ is the $Z^0$ boson pole mass and the SMEFT WCs
$C_{ee}^{ijkm}$, $C_{le}^{ijkm}$ and $C_{ll}^{ijkm}$ (where $i,j,k,m \in
\{1,2,3\}$ are family indices) are written in the Warsaw
convention~\cite{Grzadkowski:2010es}. $g_Z^{e_X}$ is the
di-$X$-handed electron coupling  
to the $Z^0$ boson including tree-level corrections from SMEFT and
$g_Z^{e_{j_Y}}$ is the di-$Y$-handed $j^{th}$-family coupling to the $Z^0$
boson~\cite{Brivio:2017vri}.
$\Gamma_Z$ is the $Z^0$ boson's total width. 
Summing over the final spins and averaging over initial spins, we obtain a
differential cross-section
\begin{eqnarray}
  \frac{d \sigma}{d t} = \frac{1}{16 \pi} \left\{|C_{le}^{1jj1}|^2 +
  \sum_{X,Y} |N^{XY}_{1jj1}(s)|^2 \left[
    \delta_{XY} \left(1 + \frac{t}{s}\right)^2 + (1 - \delta_{XY})
    \frac{t^2}{s^2}\right]\right\}. \label{differential}
\end{eqnarray}
Integrating, we obtain the total cross-section
\begin{equation}
  \sigma(e^+ e^- \rightarrow e_j^+ e_j^-) =
  \frac{s}{48 \pi} \left\{ 3 |C_{le}^{1jj1}|^2 + \sum_{X,Y} | N^{XY}_{1jj1}(s)
  |^2\right\} \label{totmumu}
  \end{equation}
and a forward cross-section minus backward cross-section
\begin{equation}
  \sigma(e^+ e^- \rightarrow e_j^+ e_j^-)_F - \sigma(e^+ e^- \rightarrow e_j^+ e_j^-)_B
  =
  \frac{s}{64 \pi} \sum_{X,Y} \left| N^{XY}_{1jj1}(s)\right|^2 (2 \delta_{XY}-1).
  \label{afbmumu}
  \end{equation}
The dimension-6 WC-SM interference terms in (\ref{totmumu}) and (\ref{afbmumu})
agree with the interference terms derived in Ref.~\cite{Falkowski:2015krw} (in the
 parameter space considered in Ref.~\cite{Falkowski:2015krw}, such
interference terms encapsulate the dominant effects of the SMEFT operators and 
were the only ones presented explicitly there). 
(\ref{differential}) shows that the part proportional to
$|C_{le}^{1jj1}|^2$ does not interfere with the rest of the matrix element due
to its different helicity-flavour structure (as already noted in
Ref.~\cite{Falkowski:2015krw}). 

In Ref.~\cite{ALEPH:2013dgf}, the SM prediction for the cross-section
is given including some higher-order
one-loop contributions. By using the \emph{ratio} of the predicted
cross-section to the SM prediction in the constraints, we can effectively
include the dominant effects of these higher order contributions. 
Our implementation
in \flavio\ therefore uses such a ratio. Both the ratios of the total
cross-section and the 
forward cross-section minus the backward cross-section are used for the
following LEP 2
centre-of-mass 
energies:
\begin{equation}
E/\text{GeV}\in \{130.3, 136.3, 161.3, 172.1, 182.7, 188.6, 191.6, 195.5, 
199.5, 201.8, 204.8, 206.5\}.
\end{equation}
Correlations between the various measurements are neglected for $e^+e^-
\rightarrow \ell^+ \ell^-$, where $\ell \in \{\mu, \tau\}$.

\subsection{LEP: Bhabha scattering}

We
calculate the tree-level polarised amplitude ${\mathcal M}$ for
$e^+(p_2) e^-(p_1) \rightarrow e^+(q_2)e^-(q_1)$ in the massless electron
approximation. 
\begin{eqnarray}
  -i{\mathcal M} &=&
  -I_t\frac{e^2}{t} \ + I_s \frac{e^2}{s} +   \sum_{X,Y} \left\{
  - \left( \bar u \gamma^\mu P_X u\right) \left(\bar v \gamma_\mu P_Y v \right) t_{XY}
  +\left( \bar u \gamma^\mu P_X v\right) \left(\bar v \gamma_\mu P_Y u \right) s_{XY}
  \right\},
\end{eqnarray}
where $u:=u(p_1)$, $v:=v(q_2)$, $\bar u:=\bar u(q_1)$ and $\bar v:=\bar v(p_2)$ are the
usual positive and negative energy 4-component Dirac 
spinors of the electron field and
\begin{eqnarray}
  I_t&:=&\left( \bar u \gamma^\mu u\right) \left(\bar v \gamma_\mu v
  \right), \nonumber \\
  I_s&:=&\left( \bar u \gamma^\mu v\right) \left(\bar v \gamma_\mu u
  \right), \nonumber\\
  t_{XY}&:=& \frac{g_{e_X}g_{e_Y}}{t - M_Z^2 + i \Gamma_Z M_Z} + \frac{C_{XY}
    + C_{YX}}{2}, \nonumber \\
  s_{XY}&:=& \frac{g_{e_X}g_{e_Y}}{s - M_Z^2 + i \Gamma_Z M_Z} + \frac{C_{XY}
    + C_{YX}}{2} 
  \end{eqnarray}
and $C_{LL}:={C_{ll}^{(1)}}^{1111}$, $C_{RR}:=C_{ee}^{1111}$,
$C_{LR}:=C_{le}^{1111}$ and $C_{RL}:=C_{le}^{1111}$. 
The spin summed/averaged differential cross-section in the centre-of-mass
frame is then
\begin{eqnarray}
  \frac{d \sigma}{d \cos \theta}& =& \frac{1}{32 \pi s} \left(
  2 e^4 \left[ \frac{u^2+s^2}{t^2} + \frac{u^2 + t^2}{s^2} +
    \frac{2u^2}{st}\right]  + \right. \label{bhabha} \\ &&
  \sum_{X,Y} \left\{
   \frac{2e^2}{t} \left(
  \text{Re}(t_{XY}) \left[ u^2 \delta_{XY} + s^2(1-\delta_{XY})\right] 
  +  \text{Re}(s_{XY})u^2 \delta_{XY}\right) +\right. \nonumber \\ &&
  \frac{2e^2}{s} \left(
   \text{Re}(s_{XY}) \left[ u^2 \delta_{XY} +
    t^2(1-\delta_{XY})\right] +
   \text{Re}(t_{XY}) u^2  \delta_{XY}\right)
   + \nonumber \\ && \left. \left.
  |t_{XY}|^2 [u^2 \delta_{XY} + s^2(1 - \delta_{XY})] +
  |s_{XY}|^2 [u^2 \delta_{XY} + t^2(1 - \delta_{XY})] +
  \right. \right. \nonumber \\ && \left. \left.
  2 \text{Re}(t_{XY}^\dag s_{XY}) u^2 \delta_{XY}
  \right\}
  \right), \nonumber
\end{eqnarray}
where $\theta$ is the scattering angle. 
Extracting the dimension-6 SMEFT WC-SM interference terms from (\ref{bhabha}),
we observe agreement 
with Ref.~\cite{Falkowski:2015krw}, providing an independent check on both
calculations. In order 
to implement the Bhabha 
scattering constraints into \flavio\ we have integrated (\ref{bhabha}) with
respect to $\cos \theta$ (utilising $t=-s(1-\cos\theta)/2$ and
$u=-s(1+\cos\theta)/2$), since the combined LEP2 data in
Ref.~\cite{ALEPH:2013dgf} are given in bins of $\cos \theta$. 
The resulting expression is rather large, so we do not list it here, although
we note that it can be found in the ancillary information stored with the {\tt
  arXiv} 
version of this paper. 

Ref.~\cite{ALEPH:2013dgf} combined LEP-experiment cross-sections for
$e^+e^-\rightarrow 
e^+e^-$ for centre-of-mass energies between 189 GeV and 207 GeV and in bins of
$\cos \theta$ in the interval $[-0.9,0.9]$ were presented. The SM prediction of
the binned 
cross-sections were also given and we shall again constrain the ratio of
the measured cross-section to the SM prediction in order to constrain SMEFT
operators.  
Here, correlations between measurements were given in Ref.~\cite{ALEPH:2013dgf}
and are taken into account. We again take ratios of each measurement with
the SM prediction
in order to effectively utilise some higher order corrections that
were included in the SM prediction; calculating their correlation
coefficients, we see that these ratios 
have identical correlations to those between the original measurements, since
the normalising factor cancels between the numerator and the denominator.

LEP-2 cross-sections and resulting constraints have been
presented and calculated specifically for a class of $Z^\prime$ models in
Ref.~\cite{Das:2021esm}. 
In the present paper, despite our application of the LEP-2 constraints to
$Z^\prime$ models, we instead find it more convenient to first match to SMEFT
and then apply the constraints in terms of said operators. There are two
reasons for this: firstly, it fits naturally into the 
\flavio\ \emph{modus operandi}, and secondly, the cross-sections and
implementation within \flavio\ could have applicability to
other models, provided that the new physics state is significantly
more massive than LEP-2 energies so the SMEFT truncation at dimension-6
remains a good approximation.

\section{Fits \label{sec:fits}}

For each of the models in the set identified in \S\ref{sec:models}, we perform
a fit of $\theta_{sb}$ and $g_{Z^\prime} / M_{Z^\prime}$
to various data by using \smelliB{}~\cite{Aebischer:2018iyb}, 
\flavioB{}~\cite{Straub:2018kue}
and \wilsonB{}~\cite{Aebischer:2018bkb} (to a good
approximation broken only by small loop corrections, the WCs -- and thus the
predictions of observables which are affected by them -- only depend upon the
ratio $g_{Z^\prime}/M_{Z^\prime}$ rather than on $g_{Z^\prime}$ and $M_{Z^\prime}$ separately). Practically, in the
numerics, we set $M_{Z^\prime}=3$ TeV throughout this paper. 
\smelli\ and
\flavio\ have been updated to include the LEP-2 measurements as described in
\S\ref{sec:LEP}, as well as the updated combination of
measurements of $B_{d,s} \rightarrow \mu^+\mu^-$ branching ratio measurements
from Ref.~\cite{Allanach:2022iod}.

\begin{figure}
  \begin{center}
    \includegraphics[width=0.8 \textwidth]{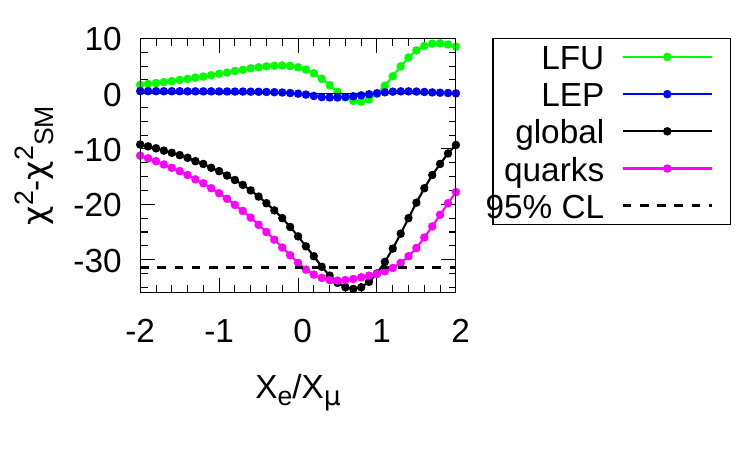}
  \end{center}
  \caption{$\chi^2$ improvement with respect to that of the SM as a
    function of electron charge divided by muon charge,
    $X_e/X_\mu$. $X_\tau=3-X_e-X_\mu$ at each point, as implied by (\ref{b3mL}). A
    negative value of $\chi^2 - \chi^2_{SM}$ indicates an 
    \emph{improvement} of the fit with respect to the SM whereas a positive
    value indicates a worse fit than the SM\@. `LFU´ contains 23 observables such
    as $R_K$ and $R_{K^\ast}$, which test lepton flavour universality. `LEP´
    contains the 148 $e^+e^-\rightarrow l^+l^-$ measurements discussed in
    \S\ref{sec:LEP}. `quarks´ contains 224 other $b \rightarrow s$ transition measurements defined
    in \flavioB. Under the hypothesis that the model line is correct, the
    region where the `global' results are  \emph{below} the marked dashed line
    is within the 95$\%$ fit region.
    \label{fig:chisq}} 
\end{figure}
We show the $\chi^2$ improvement with respect to the SM in
Fig.~\ref{fig:chisq} as a function of the $U(1)_{X}$ electron charge divided by
the muon charge, $X_e/X_\mu$.
All fit outputs that we present are approximately only sensitive to this
ratio of charges aside from the value of the best-fit gauge coupling
$g_{Z^\prime}$ and the mixing angle $\theta_{sb}$: these are sensitive to the
value of $X_\mu$ itself, and shall be presented here for the default value of
10 for this variable\footnote{Notice that because $X_\tau$
depends upon $X_\mu$ and $X_e$ separately, the predicted rates for
$b\rightarrow s \tau^+ \tau^-$ also depend upon their individual values and
not just their ratio. Since there currently is no measurement of such rates,
we do not present them here.}.  
The 23 measurements of LFU
observables prefer $0.2 \leq X_e/X_\mu\leq 1.2$ to 95$\%$ CL. The LEP-2 constraints pass through the origin, since $X_e=0$
decouples the $Z^\prime$ from electron pairs, meaning that the tree-level
predicted cross-sections are identical to those of the SM\@. On the other hand, the `quarks' set of
observables, 
which contains angular distributions of $B \rightarrow K^\star \mu^+ \mu^-$
decays and of branching ratios in different bins of di-muon invariant mass
squared, enjoys a larger effect, improving $\chi^2$ on the SM value by over 35
units in the domain taken. 
Adding all effects together in the `global' fit, we see that in fact
$0.3 < X_e/X_\mu < 1.1$
is within the 95$\%$ CL preferred region. 

\begin{figure}
  \begin{center}
    \includegraphics[width=0.47 \textwidth]{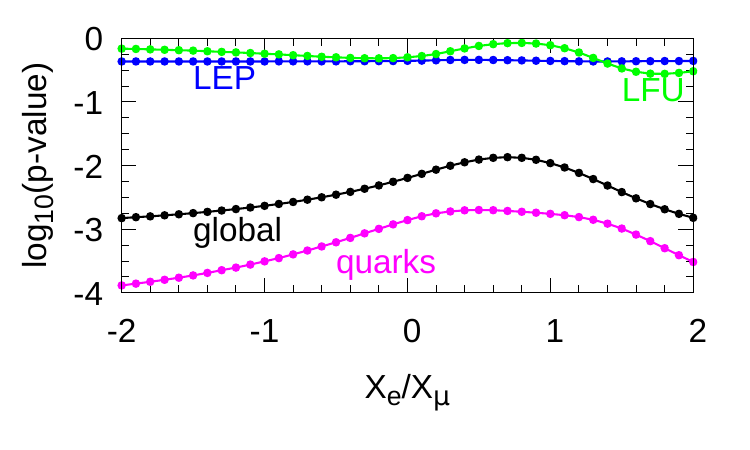}
    \includegraphics[width=0.47 \textwidth]{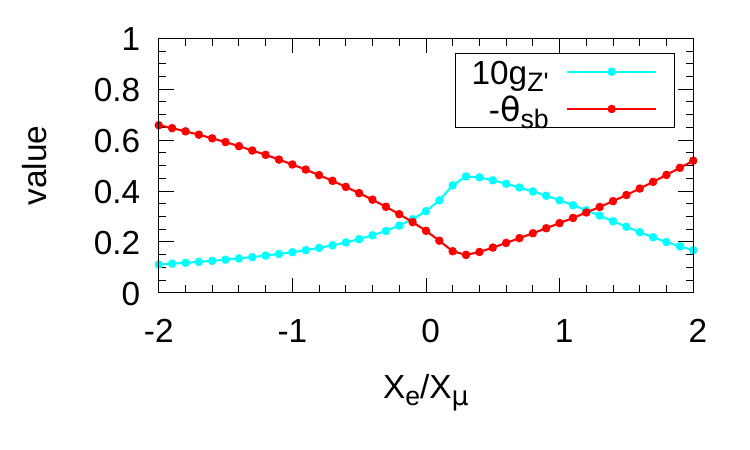}\\
        \includegraphics[width=0.47 \textwidth]{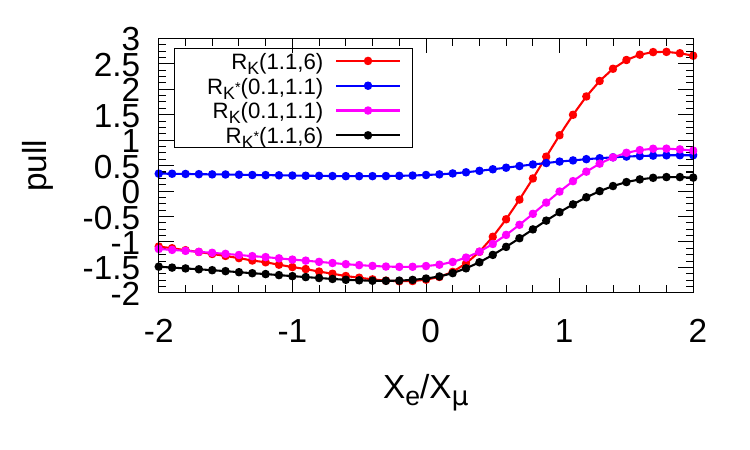}
  \end{center}
  \caption{(left panel) $p-$values, (right panel) best-fit parameters for
    $X_\mu=10$ and
    (bottom panel) 2022 LHCb LFU measurements
    associated with each model, as a  function of $X_e/X_\mu$, for $M_{Z^\prime}=3$
    TeV. $X_\tau=3-X_e-X_\mu$ at each point, as implied by (\ref{b3mL}).
    In the left panel,
    `LFU´ contains 23 observables (including the aforementioned 2022 LHCb LFU
    measurements) such 
    as $R_K$ and $R_{K^\ast}$, which test lepton flavour universality. `LEP´
    contains the 148 $e^+e^-\rightarrow l^+l^-$ measurements discussed in
    \S\ref{sec:LEP}. `quarks´ contains 224 other \bsmm\ measurements defined
    in \flavioB. In the bottom panel, the legend displays the domain of
    invariant mass squared in GeV$^2$ in parenthesis and `pull' is defined as
    $(p-e)/\sigma$, where $p$ is the theoretical prediction of the best-fit
    point of the $U(1)_X$ model, $e$ is the experimental central value and
    $\sigma$ is     the 
    experimental uncertainty, ignoring correlations with other observables.
    \label{fig:best_fit_props}} 
\end{figure}
The $p-$values associated with each fit, as well as the best-fit values of
parameters, are displayed in the left-hand panel of
Fig.~\ref{fig:best_fit_props}. 
We see that the $p-$values of the `LEP' and `LFU' categories defined are above the
{}.05 level, indicating a reasonable fit to each category. The `quarks' category is \emph{not} fit well; this could be due either to: the
\flavio\ predictions not having large enough theory errors ascribed to them,
unaccounted for experimental systematic errors
or that the set of models we have chosen is not the best one to
describe the data in the category.
Because of this, global
$p-$values show a poor fit overall, but we must bear in mind that the
SM is a much worse fit. 
In the right-hand panel of Fig.~\ref{fig:best_fit_props}, we see
some trends. The fact that $g_{Z^\prime}$ falls towards the right-hand side and
left-hand side of the plot can be seen as due to the fact that LEP constraints
will prefer a smaller value of $g_{Z^\prime}$ when $|X_e|$ is large, since then the
$Z^\prime$ coupling to electrons is higher. When $g_{Z^\prime}$ is smaller,
$\theta_{sb}$ is higher. This is expected: since the non-zero tree-level
new physics WET WC is constrained by the model to be
\begin{equation}
  C_9^{(\mu)} = -\frac{X_\mu {Q_3}}{2} \frac{g_{Z^\prime}^2}{M_{Z^\prime}^2}
  \sin 2 \theta_{sb}.\label{C9}
\end{equation}
  Requiring some particular fixed value of $C_9^{(\mu)} \neq 0$ to fit
the `quarks' category of observable, we see that $g_{Z^\prime}/M_{Z^\prime}$ would tend to move in the opposite
direction to $\theta_{sb}$. 

The left-hand panel of Fig.~\ref{fig:best_fit_props} confirms that out of our
set, globally, $X_e/X_\mu\approx 1/2$ provides a reasonable fit to the
experimental measurements included.
The optimal model at this value of the ratio corresponds to fermionic
charges of 
$3B_3 - 7L_e - 10L_\mu + 14 L_\tau$.
Some may feel that, aesthetically, some of the charges in this assignment are
rather large. 
This suggests that we investigate a different 
model in our set with a similar ratio of $X_e/X_\mu$ but 
with smaller $X_\mu$. By adjusting $g_{Z^\prime}$, we may expect the fit then
to reach a similar $\chi^2-\chi^2_{SM}$ for the LEP observables.
We also expect that $\theta_{sb}$
will then change to keep the value of (\ref{C9}) invariant. There should be
small corrections to this overall $\chi^2-$invariant
picture from
$\Delta m_s$ the variable inferred from $B_s-\overline{B_s}$ mixing, which has
a different dependence upon $g_{Z^\prime}$ and $\theta_{sb}$ to $C_9$.
There are also some negligible corrections coming from 
the different gauge coupling affecting the renormalisation
between $M_{Z^\prime}$ and $M_Z$.
One charge assignment with $X_e/X_\mu=1/2$ (i.e.\ within the 95$\%$ CL
constraint along the model line) which is anomaly-free
is\footnote{For this particular choice, $X_\tau=0$, leading to no tree-level
new physics contribution to $b \rightarrow s \tau^+ \tau^-$
transitions. However, 
we note that generically $X_\tau \neq 0$ as other choices of $X_e$ and $X_\mu$
show.} 
 $3B_3 - L_e - 2 L_\mu$.
We shall investigate this model in more detail now. 

\subsection{$3B_3-L_e-2L_\mu$ \label{sec:mod}}

Here, we perform a new fit to the $3B_3-L_e-2L_\mu$ model; the result is 
displayed in Table~\ref{tab:nice_model}. 
\begin{table}
  \begin{center}
    \begin{tabular}{|c|cc||lc|} \hline
             & $\chi^2-\chi^2_{SM}$ & $p-$value & measurement             & pull\\ \hline
      LFU    &   -0.48              & 0.75       & $R_{K^\ast}[0.1,\ 1.1]$ & 0.4\\
      LEP    &    0.12              & 0.45      & $R_{K^\ast}[1.1,\ 6]$     & -1.3\\
      quarks &   -34.1             & 2.3$\times 10^{-3}$       & $R_K[0.1,\ 1.1]$      & -1.1\\ 
      global &   -33.8             & 0.014      & $R_K[1.1,\ 6]$          & -0.9 \\ \hline
    \end{tabular}
  \end{center}
  \caption{Quality-of-fit for the $3B_3-L_e-2L_\mu$ $Z^\prime$ model and the
    pulls of the 
    2022 LHCb LFU measurements. The numbers in square parenthesis refer to the
    end-points of the domain of the relevant bin of di-lepton invariant mass
    squared, in units of GeV$^2$.
    For $M_{Z^\prime}=3$ TeV, the best-fit parameters are $g_{Z^\prime}=0.206$,
    $\theta_{sb}=-0.0406$.
    \label{tab:nice_model}}
\end{table}
We see that the overall fit is of poor quality, with a
$p-$value of {}.014. The higher invariant mass-squared bins of $R_K$,
$R_{K^\ast}$ are more-or-less compatible with their experimental values (the pull are 
-0.9$\sigma$, -1.3$\sigma$, respectively).
The $B_3-L_e-2L_\mu$ model has a $\chi^2$ improvement of $33.8$ as compared to
the SM, for two additional fitted parameters. The $p-$value of the SM being
statistically as good a fit to the data\footnote{The SM is equivalent to the
parameter choice $g_{Z^\prime}=0$, $\theta_{sb}=0$ of the $B_3-L_e-2L_\mu$
model under suitable caveats; therefore we use the optimal
likelihood ratio test for two degrees of freedom to calculate the $p-$value.}
as the 
$B_3-L_e-2L_\mu$ model is $4.6 \times 10^{-8}$.  

\begin{figure}
  \begin{center}
    \includegraphics[width=0.47 \textwidth]{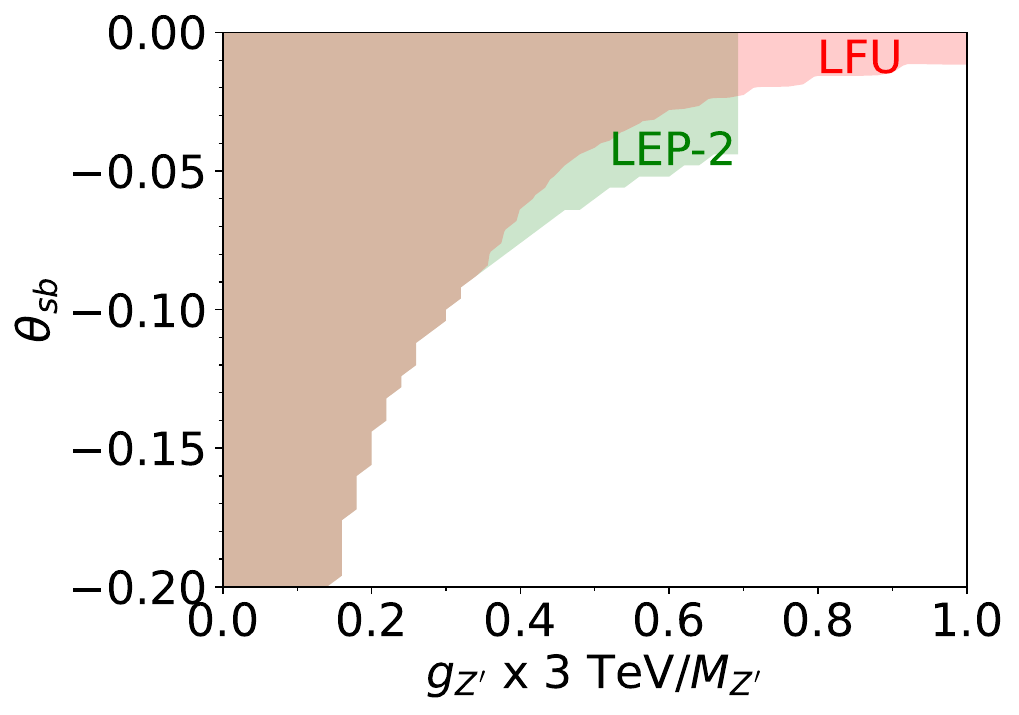}    
    \includegraphics[width=0.47 \textwidth]{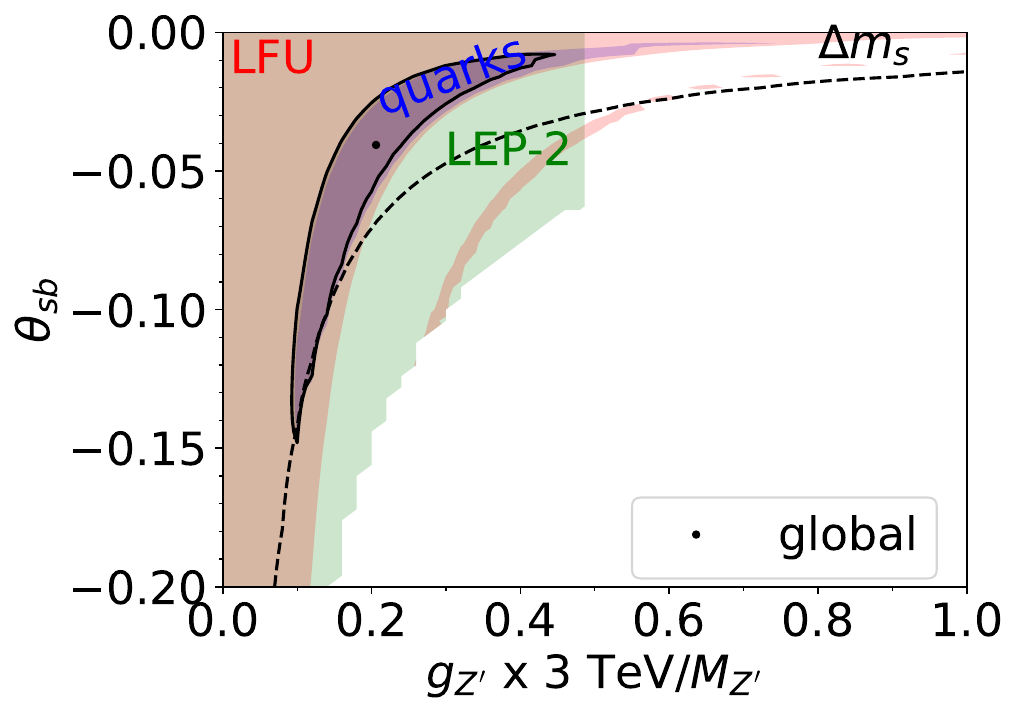}
  \end{center}
  \caption{\label{fig:param_space} Parameter space of the $3B_3-L_e-2L_\mu$
    model: (left) compatibility of different sets of observables and (right)
    constraints. `LFU´ contains 23 observables (including the
    aforementioned 2022 LHCb LFU  measurements) such 
    as $R_K$ and $R_{K^\ast}$, which test lepton flavour universality. `LEP-2´
    contains the 148 $e^+e^-\rightarrow l^+l^-$ measurements discussed in
    \S\ref{sec:LEP}. `quarks´ contains 224 other \bsmm\ measurements defined
    in \flavioB.
    In the left-hand panel, the coloured regions show where the $p-$value of
    each labelled constraint is greater than 0.05. In the right-hand panel,
    the coloured regions are the 95$\%$ confidence limit (CL) \emph{allowed}
regions defined by $\chi^2-\chi^2(\text{min})=5.99$~\cite{James:2004xla} as
shown for the labelled category of observable. The black line
encloses the 95$\%$ CL global region and the black dot gives the locus of the
best-fit point. The
region above the dashed line is compatible with the $B_s-\overline{B_s}$ mixing
constraint at the 95$\%$ CL (note that this measurement is a member of the
`quarks' set of observables). }
\end{figure}

We display the parameter space of the model in Fig.~\ref{fig:param_space}.
One should interpret the left-hand panel as testing the joint compatibility of
measurements between different categories of observable in the model. We see
that nowhere does the `quarks' category of $p-$value exceed {}.05.
However,
the LFU and LEP2 constraints are compatible in a significant part of parameter
space. 
The right-hand panel should be interpreted as parameter constraints upon the
model, \emph{assuming that the $3B_3-L_e-2L_\mu$ model hypothesis is correct}.
Here, we see that the constraints from the `quarks' category of
observable is more-or-less compatible with the LFU constraints. The LEP-2
constraints cut off the global fit contour at the top-right hand side, for
larger $g_{Z^\prime}/M_{Z^\prime}$. The $B_s-\overline{B_s}$ mixing
constraint cuts off the global-fit region at the lower left-hand side.
A curved region at the bottom right-hand side of the plot has too large
flavour changing effects in general for \flavio\ to return a numerical answer,
which explains why some of the constraints (notably from LEP-2) are bounded
there. Such regions are highly ruled out by flavour
measurements anyway. 
From the left-most plot, 
we conclude that the model does not describe the `quarks' category of data as
predicted by \flavio\ well.
This could be due to: problems in experimental
measurements or problems in \flavio{}'s theoretical predictions or indeed the
model being incorrect (although not nearly as incorrect as the SM!)

\begin{figure}
  \begin{center}
    \includegraphics[width=0.5 \textwidth]{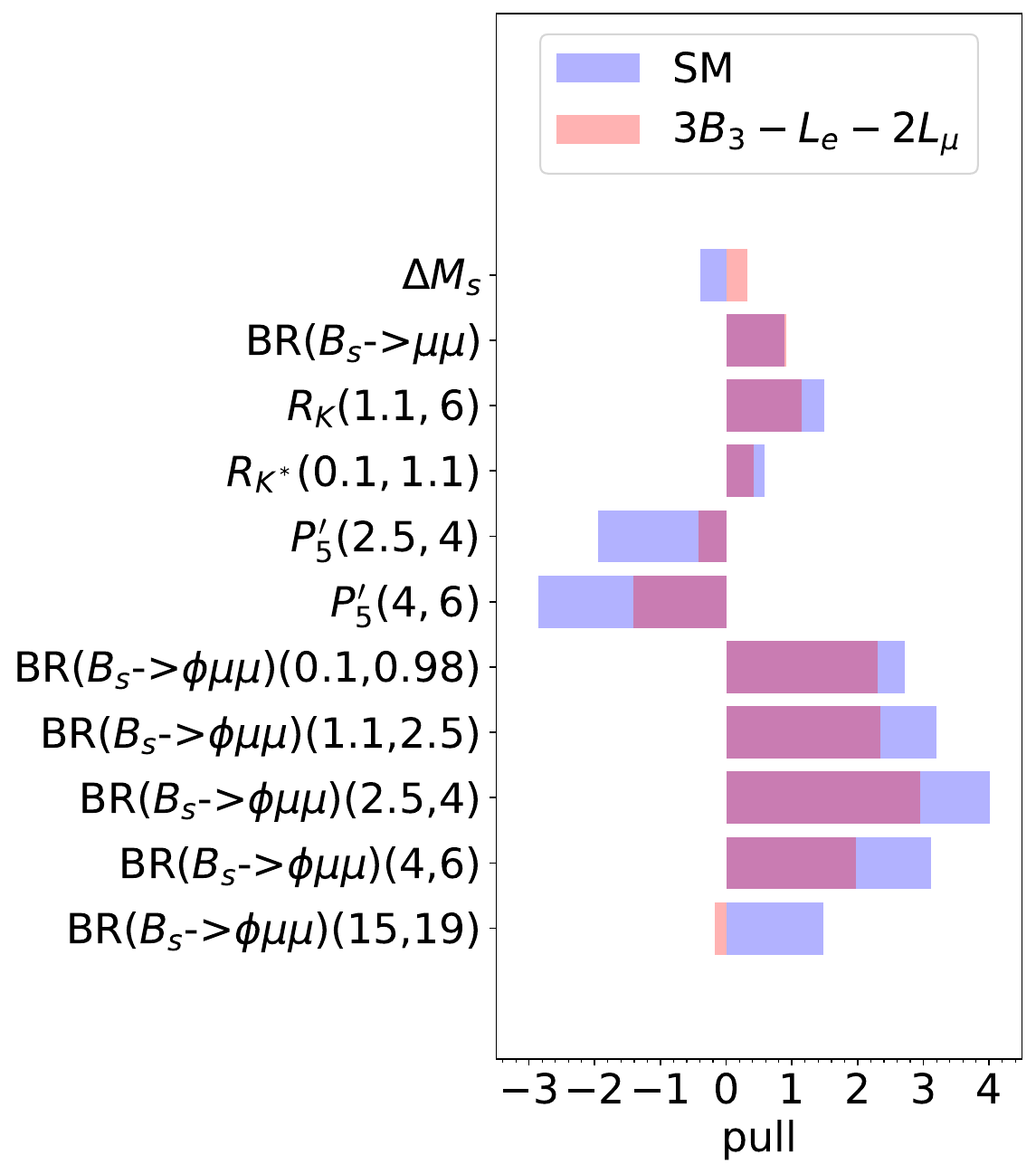}
  \end{center}
  \caption{\label{fig:obs} Various pulls of interest for the SM and the
    best-fit point of 
  the $3B_3-L_e-2L_\mu$ model. Pull is defined as theory prediction minus the
  experimental central value divided by uncertainty. Correlation between
  observables is neglected in the calculation of the pull.}
\end{figure}
We show the pulls of some observables of interest in Fig.~\ref{fig:obs}. We
see that although the $3B_3-L_e-2L_\mu$ model fits $B_s-\overline{B_s}$ mixing (as measured by $\Delta m_s$),
and $R_{K^\ast}(1.1, 6)$ less well than the SM,
it ameliorates the fit to more of the other observables. 
Various bins of $BR(B_s \rightarrow \phi \mu^+\mu^-)$, whilst fitting
\emph{better} than in the SM, are still far from optimal (the egregious one
being 3$\sigma$ between 2.5 and 4 GeV$^2$ in di-muon invariant mass squared).
This goes some way to confirming the assertion
in the discussion of Fig.~\ref{fig:best_fit_props}: that the fit to the
`quarks' category of observable is far from perfect.

\section{Conclusion \label{sec:conc}}

We have critically re-examined $Z^\prime$ models that can
significantly ameliorate the $b \rightarrow s \mu^+ \mu^-$ anomalies in global
fits. The 2022 re-analysis of the $R_{K}$ and $R_{K^\ast}$ observables by
LHCb implies that, if the $b \rightarrow s \mu^+ \mu^-$ anomalies are due to
beyond the SM effects, there may well also be beyond the SM effects in
$b \rightarrow s e^+ e^-$. One possible explanation is that of a TeV-scale
$Z^\prime$ boson 
that couples dominantly to third family quarks, to $s \bar b$ and $b \bar s$
through weak mixing effects, to di-muon pairs and to di-electron pairs in
addition. We identified a one-rational-parameter family of models
which, in the first two-family charged-lepton sector, 
interpolates between a $Z^\prime$ only coupling to di-muon pairs
and a $Z^\prime$ which couples 
to di-electron pairs and to di-muon pairs with equal strength.
Here, the coupling strength is directly proportional to the $U(1)_{X}$ charge of
the leptonic field in question. 
By coupling a $Z^\prime$ to di-electron pairs, one obtains
constraints from LEP-2, which measured the scattering of $e^+e^-$ to di-lepton
pairs and observed no significant deviations from SM predictions.
One hopefully useful side-product of the present paper was to re-calculate such
predicted 
deviations resulting from relevant dimension-6 SMEFT operators. A previous
presentation in the literature~\cite{Falkowski:2015krw} has thus received an independent check.
Our calculation is presented in a more complete form than in Ref.~\cite{Falkowski:2015krw}, which guarantees that
the resulting predicted LEP-2 cross-sections are positive even in extreme
parts of parameter 
space. The calculations in the more
complete form have been programmed into {\tt smelli} and {\tt flavio}, and are
thus 
publicly available for use. In a different analysis,
other experimental data, such as electroweak
precision observables, can be varied (or indeed re-fit) and the LEP-2
constraints will change accordingly in the calculation.

One particular model, $3B_3-L_e-L_\mu-L_\tau$, was already examined in
Ref.~\cite{Greljo:2022jac} (using the numerical results from a fit to 2015
electroweak and LEP-2 data)
where it was found that there is no parameter region where each set of
experimental 
constraints is satisfied to within 1$\sigma$. We think this condition to be
overly 
restrictive 
and we prefer a global fit strategy. By widening the model space to allow
different electron and muon $U(1)_{X}$ charges, we also see how much the
preference is for an equal coupling of the $Z^\prime$ to di-electron pairs and
to di-muon pairs, as 
compared to some other ratio between the two. This main information is
displayed in Fig.~\ref{fig:chisq}. From the figure, it appears that 
zero coupling of the $Z^\prime$ boson to di-electron pairs is
a poor fit whereas 
an equal coupling to di-muon pairs
is better. We also see from the figure that a 
$Z^\prime$ which couples to di-electron pairs with about three-quarters the
strength 
with which it couples to di-muon pairs is a close-to-optimal global fit, at
least 
along one particular line in the space of rational anomaly-free charge
assignments. We 
proposed the 
$3B_3-L_e-2L_\mu$ model\footnote{We urge the reader to note that other models
involving 
$Z^\prime$ coupling to di-tau pairs also provide close-to-optimal fits, for
different values than $X_e=1$, $X_\mu=2$.}; its
properties as regards flavour changing variables 
are investigated in \S\ref{sec:mod}. Here, Fig.~\ref{fig:best_fit_props} shows a sub-optimal fit
to some observables in the `quark' category of observable
. Further 
lepton-flavour universal corrections to $C_9$ WCs coming from non-perturbative
corrections may potentially ameliorate these. 
In fact, the better fit (than the SM) in the `quarks' category comes from
the new physics contribution to $C_9$, as shown in \S\ref{sec:SMEFTop},
which the $3B_3-L_e-2L_\mu$ model does admirably. 
In a fit to 12-dimensional WET operator space~\cite{Hurth:2023jwr}, it was
shown how only $C_9$ 
significantly improves the fit: by 6.3$\sigma$
(a new physics contribution to $C_7$ improves
it by 1.7$\sigma$ and the other operators improve only by a small amount).
Thus there is no evidence to suggest that a different model would yield
significantly better gains, in the current situation\footnote{For more modest
gains, one could try
$3B_3-2L_e-3L_\mu+2L_\tau$ or $3B_3-3L_e-4L_\mu+4L_\tau$ instead as example of
close-to-optimal anomaly-free models.}.

For more generic models than this close-to-optimal model, the third generation
of leptons have a non-zero $U(1)_X$ charge of $X_\mu+X_e-3$ as implied
by anomaly 
cancellation (\ref{b3mL}). This will have potentially important
phenomenological implications, implying a non-zero tree-level new physics
contribution to rates for 
$b\rightarrow s \tau^+ \tau^-$ and
affecting the prediction of $b \rightarrow s \bar \nu \nu$.

\section*{Acknowledgements}
This work was partially supported by STFC HEP Consolidated grant
ST/T000694/1. We thank the Cambridge Pheno Working Group for discussions
(especially B~Capdevila-Soler for a detailed comparison with
Ref.~\cite{Alguero:2023jeh} and E~Loisa for pointing out a bug in the LEP
di-lepton constraint) and P~Stangl for helpful advice on the
calculation of the  
LEP2 constraints and their insertion into the {\tt flavio} and {\tt smelli}
computer programs. We thank CERN for hospitality extended while this work was
carried out.

\bibliographystyle{JHEP-2}
\bibliography{dy}

\end{document}